\documentclass[a4paper,12pt]{article}
\pdfoutput=1 
\usepackage{jheppub} 
\usepackage[T1]{fontenc} 
\usepackage{epsfig}
\usepackage{float}
\def\beq{\begin{equation}}
\def\eeq{\end{equation}}
\def\beqn{\begin{eqnarray}}
\def\eeqn{\end{eqnarray}}
\def\ee{e^+e^-}
\def\as{\alpha_{\rm S}}
\def\bas{\overline{\alpha}_{\rm S}}

\def\e{{\rm e}}

\def\k{\kappa}
\def\l{\lambda}

\def\xj{\xi_j}

\def\nn{\nonumber\\}

\title{Associated jet and subjet rates in light-quark and gluon jet discrimination}

\author[a]{Biplob Bhattacherjee,}
\author[b]{Satyanarayan Mukhopadhyay,}
\author[b,c]{Mihoko M. Nojiri,}
\author[c]{Yasuhito Sakaki}
\author[d]{and Bryan R. Webber}

\affiliation[a]{Centre for High Energy Physics, Indian Institute of Science, Bangalore, India}
\affiliation[b]{Kavli IPMU (WPI), The University of Tokyo, Kashiwa, Chiba 277-8583, Japan}
\affiliation[c]{KEK Theory Center and Sokendai, Tsukuba, Ibaraki 305-0801, Japan}
\affiliation[d]{Cavendish Laboratory, J.J. Thomson Avenue, Cambridge, UK}

\emailAdd{biplob@cts.iisc.ernet.in}
\emailAdd{satya.mukho@ipmu.jp}
\emailAdd{nojiri@post.kek.jp}
\emailAdd{sakakiy@post.kek.jp}
\emailAdd{webber@hep.phy.cam.ac.uk}

\preprint{\\Cavendish-HEP-15/01\\KEK-TH-1790\\IPMU15-0008}

\abstract{We show that in studies of light quark- and gluon-initiated jet discrimination, it is important to include the information on softer reconstructed jets (associated jets) around a primary hard jet. This is particularly relevant while adopting a small radius parameter for reconstructing hadronic jets. The probability of having an associated jet as a function of the primary jet transverse momentum ($p_T$) and radius, the minimum associated jet $p_T$ and the association radius is computed up to next-to-double logarithmic accuracy (NDLA), and the predictions are compared with results from {\tt Herwig++}, {\tt Pythia6} and {\tt Pythia8} Monte Carlos (MC). We demonstrate the improvement in quark-gluon discrimination on using the associated jet rate variable with the help of a multivariate analysis. The associated jet rates are found to be only mildly sensitive to the choice of parton shower and hadronization algorithms, as well as to the effects of initial state radiation and underlying event. In addition, the number of $k_t$ subjets of an anti-$k_t$ jet is found to be an observable that leads to a rather uniform prediction across different MC's, broadly being in agreement with predictions in NDLA, as compared to the often used number of charged tracks observable.}

\begin{document} 
\maketitle
\flushbottom
\section{Introduction}
Hadronic jets are the most abundant objects at a proton-proton collider like the LHC, and it is a major challenge to separate the signals being looked for from standard model (SM) backgrounds in multijet final states. One promising direction that has recently received attention in both theoretical and experimental studies is that the separation of light quark-initiated jets from gluon-initiated ones can be viable in these search channels.
Quarks are often encountered in the decays of new particles predicted by scenarios beyond the standard model, as well as in the decay of the weak bosons, Higgs and top quark in the SM itself. On the other hand, in the corresponding SM backgrounds involving multiple hard jets, there is a larger fraction of gluon-initiated jets from QCD radiation. Here, quark- or gluon-initiated jets (henceforth simply referred to as quark and gluon jets) refer to the parton in the hard process at leading order in perturbation theory that initiates the parton shower. Based on the difference in the radiation pattern of quarks and gluons, a likelihood based discriminant can be built to separate decay jets from QCD radiation jets with a certain efficiency~\cite{Schwartz1}. 

Several variables have been proposed to separate quark and gluon jets, mostly relying on the fact that a gluon of similar energy leads to more soft emissions compared to a quark. This includes both discrete variables like the number of charged tracks inside the jet cone, as well as continuous ones like the width of a jet and energy-energy-correlation (EEC) angularity~\cite{Schwartz1,Schwartz2,Schwartz3,Larkoski1,Larkoski2}. ATLAS and CMS collaborations have also studied the discrimination of light quarks from gluons along these lines with the 7 and 8 TeV LHC data respectively~\cite{ATLAS,CMS}. Using data samples with "enriched quark and gluon content", data-based taggers were also developed, and compared to the predictions from Monte Carlo (MC) simulations. While there are differences between the predictions of different MC's, as well as between the data-based tagger and the MC results, they are consistent with each other within the large systematic uncertainties at present.

An important question in this regard is the proper choice of a jet algorithm and radius parameter. In the LHC environment, in order to keep the contribution of the underlying event and multiple proton-proton collisions at a minimum, for multijet processes the standard choice is an anti-$k_t$ algorithm with radius parameter $R=0.4$. In addition, in the ATLAS study mentioned above, jets are required to satisfy an isolation criterion: a jet is considered isolated if there is no other reconstructed jet within a cone of size $\Delta R < 0.7$ (where $\Delta R = \sqrt{(\Delta \eta)^2+(\Delta \phi)^2}$ is the standard distance measure in the pseudorapidity-azimuthal angle plane). An optimum choice for the jet radius parameter was discussed in Refs.~\cite{Dasgupta,Salam:2009jx} for quark and gluon jets as a function of their transverse momenta ($p_T$), and it was observed that one usually requires a larger radius for a gluon jet in order for the parton $p_T$ to be close to the jet $p_T$. However,  for experimental purposes it is advantageous to use a fixed and small radius parameter for the jets, for reasons mentioned above. Therefore, we propose to recover the missed information on radiation from the parent parton outside the chosen jet radius by including softer reconstructed jets that can be present (with a calculable probability) around a certain radius of a primary hard jet. These softer jets are referred to as "associated jets" in this study. It is important to note here that imposing an isolation criterion as above while studying quark and gluon jet properties might not be appropriate, since it leads to rejecting a fraction of the jet candidates beforehand, and thus biasing the sample to ones where the initial quark or gluon has not radiated outside the adopted jet radius.

We first compute the associated jet rates in QCD to next-to-double logarithmic accuracy in Sec.~\ref{sec:2}, and then compare the analytical results with those from different parton shower MC's in Sec.~\ref{sec:3}. Using the information on the presence (or absence) of associated jets can improve the discrimination of quarks and gluons. We demonstrate this through a multivariate analysis in Sec.~\ref{sec:4}. Several combinations of jet discrimination variables are tried out, and an attempt is made to determine an optimum choice. Even though we include standard discrimination variables like the number of charged tracks as inputs to our multivariate analysis, it should be emphasized that they are subject to MC ambiguities stemming from parton shower algorithms and their associated parameters, and tunings of hadronization and underlying event (UE) models. However, in order to judge the improvement in tagger performance on using the associated jet rates, we compare the performance of different sets of variables within the same MC. 

In Secs.~\ref{sec:5} and~\ref{sec:6} we study the use of the number of subjets of a jet (defined with an exclusive $k_t$ algorithm) in place of the number of charged tracks, since the different MC prediction tend to be similar for the former observable. We compute the subjet rates upto NDLA as well, and compare the NDLA results with predictions from different MC's. Our results on both associated jets and subjets are summarized in Sec.~\ref{sec:7}. We discuss the 2-dimensional joint distributions of the three discrimination variables used as inputs in the multivariate analysis in an Appendix.

\section{Associated jet rates: analytical calculations}
\label{sec:2}
To begin with, let us define the longitudinally invariant jet algorithms~\cite{,Catani:1991hj,Catani:1993hr,Ellis:1993tq,Cacciari:2008gp} adopted in this study.
The distance measures between each pair of objects $i$ and $j$ ($d_{ij}$), and between an object and the beam ($d_{iB}$) are given by
\begin{align}
d_{ij} =& \min\{p_{ti}^{2p},p_{tj}^{2p}\}\frac{\Delta R_{ij}^2}{R^2}\;,\nn
d_{iB} =& p_{ti}^{2p}\;,
\label{eq:genktpp}
\end{align}
where $p_{ti}$, $y_i$ and $\phi_i$ are the transverse momentum, rapidity and azimuth of object $i$, respectively, $\Delta R_{ij}^2  \equiv (y_i-y_j)^2+(\phi_i-\phi_j)^2$, and $R$ is the jet radius parameter. The jet algorithm in use is fixed by the parameter $p$, with $p=1,0,-1$ for the $k_t$~\cite{Catani:1993hr}, Cambridge/Aachen~\cite{CamOrig,CamWobisch} and anti-$k_t$~\cite{Cacciari:2008gp} algorithms, respectively. At any stage of clustering, if a $d_{ij}$ is the smallest measure we
combine objects $i$ and $j$.  If $d_{iB}$ is the smallest we call $i$ a jet and remove it from the clustering list.  This procedure is continued until there are no more objects left to cluster. 

\begin{figure}[]
\centering 
\includegraphics[width=0.8\textwidth]{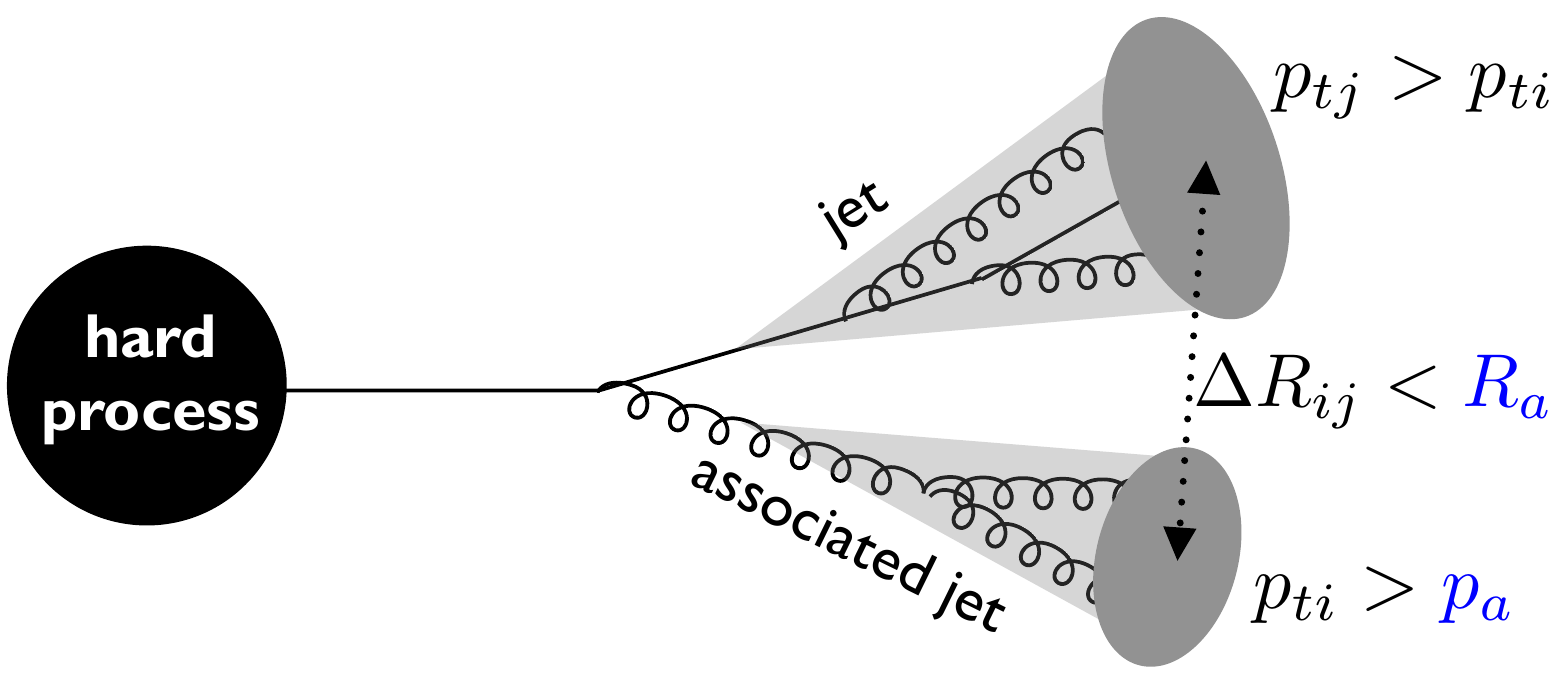}
\caption{\label{fig:concept} A schematic illustration of associated jets, and the relevant variables which determine the associated jet rate (see text for details).}
\end{figure}

Once a primary jet $j$ has been defined, say using the anti-$k_t$ algorithm with radius parameter $R$,  we define a nearby jet $i$ with $p_{tj}>p_{ti}>p_a$ and $R<\Delta R_{ij}<R_a$ as an {\it associated jet}.  Thus the associated jet rates are functions of the primary jet $p_t=p_j$, its radius $R$, the association radius $R_a$ and the minimum associated jet $p_t=p_a$. In Fig.~\ref{fig:concept} we illustrate the idea of an associated jet schematically, and show the relevant variables that determine the associated jet rate.

In perturbative QCD, the rate of $n$-jet production from a primary object of type $i$ ($i=q,g$ in this case), $R_n^i$, can be obtained from the associated generating function~\cite{Konishi:1979cb,Dokshitzer:1991wu,Ellis:1991qj,Gerwick:2012fw}
\beq
\Phi_i(u) = \sum_n R_n^i u^n\,.
\eeq
We can recover the jet rates by differentiating at $u=0$, 
\beq
R_n^i = \frac 1{n!}\left.\frac{d^n\Phi_i}{du^n}\right|_{u=0}\,.
\eeq

The jet rates $R_n^i=R_n^i(p_j,\xi)$ are functions of the trigger jet transverse momentum $p_j$, and the evolution scale for parton showering, which, for hadron-hadron collisions is taken as $\xi=\Delta R^2/2$. This is equivalent to the evolution scale for coherent parton showering, $\xi\equiv 1-\cos\theta$, with $\theta$ being the emission angle ($\Delta R^2/2 \approx\theta^2/2 \approx 1-\cos\theta$). To be resolved, an emission must have $\xi>\xj=R^2/2$ and $p_t>p_a$. Since the jet rates $R_n^i$ include the trigger jet $j$, the probability of $n$ {\em associated jets} for a jet of type $i$ with transverse momentum $p_j$ is
\beq
P_n^i = R_{n+1}^i(p_j,\xi_a)\;.
\eeq
Here, $\xi_a=R_a^2/2$, with $R_a$ being the association radius defined above. 

The generating functions $\Phi_i(u)$ were computed in the context of $\ee$ collisions in Ref.~\cite{Gerwick:2012fw}, upto next-to-double logarithmic accuracy (NDLA). Here,  leading double and next-to-double logarithms refer to $\as^n\log^{2n}$ and $\as^n\log^{2n-1}$, where the logarithms are those of $R_a/R$ and/or $p_j/p_a$. For $p_a$ sufficiently large, these terms are determined by the timelike showering of final-state partons, while contributions from initial-state showers and the underlying event can be avoided. Following the same methods as in Ref.~\cite{Gerwick:2012fw} for hadron hadron collisions, for $\xi>\xj$ and $p_j>p_a$, we have the quark and gluon generating functions to NDLA
\beqn
\Phi_q(u,p_j,\xi) &=& u +\int_{\xj}^{\xi} \frac{d\xi'}{\xi'}
\int_{p_a/p_j}^1 dz\frac{\as(k_t^2)}{2\pi}P_{gq}(z)\Phi_q(u,p_j,\xi')
\left[\Phi_g(u,zp_j,\xi')-1\right]\;,\nn
\Phi_g(u,p_j,\xi) &=& u +\int_{\xj}^{\xi} \frac{d\xi'}{\xi'}
\int_{p_a/p_j}^1dz\frac{\as(k_t^2)}{2\pi}\bigl\{P_{gg}(z)\Phi_g(u,p_j,\xi')
\left[\Phi_g(u,zp_j,\xi')-1\right]\nn
&&+ P_{qg}(z)\left[\{\Phi_q(u,p_j,\xi')\}^2-\Phi_g(u,p_j,\xi')\right]\bigr\}\;.
\eeqn
Here, the running coupling is evaluated at the transverse momentum scale of the emission, $k_t^2=z^2p_j^2\xi'$. Defining $\bas = \as(p_j^2\xi)/\pi$, i.e.\ in terms of the coupling at the hard scale, we have to NDLA
\beq\label{eq:askt}
\frac{\as(k_t^2)}{\pi}  = \bas -b_0\bas^2\left[2\ln z+\ln\left(\frac{\xi'}{\xi}\right)\right],
\eeq
with $b_0=(11C_A-2n_f)/12$.

The solution for the quark generating function is easily seen to be
\beq
\Phi_q(u,p_j,\xi) = u\exp\left\{\int_{\xj}^{\xi} \frac{d\xi'}{\xi'}
\int_{p_a/p_j}^1 dz\frac{\as(k_t^2)}{2\pi}P_{gq}(z)
\left[\Phi_g(u,zp_j,\xi')-1\right]\right\}\;.
\eeq
We can solve for the gluon generating function by iteration, and then substitute in this equation to get the complete solution.  For brevity we define the following logarithms:
\beqn\label{eq:kldef}
\k &=&\ln(p_j/p_a)\;,\;\;\;\k'=\ln(zp_j/p_a)\;,\nn
\l &=&\ln(\xi_a/\xj)=2\ln(R_a/R)\;,\;\;\;\l'=\ln(\xi'/\xj)\;.
\eeqn
In terms of these variables the NDLA quark generating
function is
\beq\label{eq:Phiq}
\Phi_q(u,\k,\l) 
  = u\exp\left\{\int_{0}^{\l} {d\l'}\int_{0}^{\k} 
d\k' \,\Gamma_q\left(\k',\l',\k,\l\right)
\left[\Phi_g(u,\k',\l')-1\right]\right\}
\eeq
where, including the full $P_{gq}$ splitting function,\footnote{We
  keep terms that are formally power-suppressed in order to
satisfy the boundary condition $P_0=1$ when $p_a=p_j$.}
\beq
\Gamma_q\left(\k',\l',\k,\l\right) =C_F\bas\left[1-\e^{\k'-\k}
+\frac 12\e^{2(\k'-\k)}\right]-C_F b_0\bas^2
\left[2(\k'-\k)+\l'-\l\right]\,.
\eeq
Defining similarly\footnote{We drop the $\bas^2$ term in $\Gamma_f$ as
  it is beyond NDLA and does not affect the boundary condition.} 
\beqn
\Gamma_g(\k',\l',\k,\l) &=&  C_A\bas\left[1-\e^{\k'-\k}
+\frac 12\e^{2(\k'-\k)}-\frac 12\e^{3(\k'-\k)}\right]-C_A b_0\bas^2
\left[2(\k'-\k)+\l'-\l\right]\,,\nn
\Gamma_f(\k',\k) &=&  \frac{n_f}4 \bas\left[\e^{\k'-\k}
-2\e^{2(\k'-\k)}+2\e^{3(\k'-\k)}\right]\,,
\eeqn
we solve the gluon generating function by iteration to second order in
$u$, which gives the probabilities for 0 or 1 associated jets:
\beqn
\Phi_{g} (u,\k, \l) &=& u \Delta_g(\k, \l) \biggl\{1+ u 
\int_0^{\l} d \l' \int_0^{\k} d\k'  
\,\bigl[\Gamma_g(\k',\l',\k,\l) \,\Delta_g(\k', \l') \nn
&&+ \Gamma_f(\k',\k)\Delta_f(\k, \l')\bigr]
+{\cal O}(u^2)\biggr\}\,,
\eeqn
where  $\Delta_q(\k,\l)$ and $\Delta_g(\k,\l)$ are the quark and gluon
Sudakov factors (the probabilities for no associated jets) and we have
defined    $\Delta_f(\k,\l)=\Delta^2_q(\k,\l)/\Delta_g(\k,\l)$.
Hence
\beqn
P_0^q &=&\Delta_q(\k,\l) 
=\exp\left\{-\int_0^{\l} d \l' \int_0^{\k} d\k' \,
\Gamma_q(\k',\l',\k,\l)\right\}\nn
&=&\exp\biggl\{-C_F\bas\l
\left[\k-\frac 34+\e^{-\k}-\frac 14\e^{-2\k}\right]
-C_F b_0\bas^2\k\l\left[\k+\frac 12 \l\right]\biggr\}\,,\\
P_0^g &=&\Delta_g(\k,\l) 
=\exp\left\{-\int_0^{\l} d \l' \int_0^{\k} d\k' 
\left[\Gamma_g(\k',\l',\k,\l)+ \Gamma_f(\k',\k)\right]\right\}\nn
&=&\exp\biggl\{-C_A\bas\l
\left[\k-\frac{11}{12}+\e^{-\k}-\frac 14\e^{-2\k}
+\frac 16\e^{-3\k}\right]\nn
&&-\frac{n_f}4\bas\l\left[\frac 23-\e^{-\k}+\e^{-2\k}
-\frac 23\e^{-3\k}\right]
-C_A b_0\bas^2\k\l\left[\k+\frac 12 \l\right]\biggr\}\,,\\
P_1^q &=&\Delta_q(\k, \l) \int_0^{\l} d \l' \int_0^{\k} d\k'  
\,\Gamma_q(\k',\l',\k,\l) \,\Delta_g(\k', \l') \\
P_1^g &=&\Delta_g(\k, \l) \int_0^{\l} d \l' \int_0^{\k} d\k'  
\bigl[\Gamma_g(\k',\l',\k,\l)\Delta_g(\k', \l')
+ \Gamma_f(\k',\k)\Delta_f(\k, \l')\bigr].
\eeqn
\section{Associated jet rates: comparison with Monte Carlo}
\label{sec:3}
We are now in a position to compare the NDLA predictions for associated jet rates discussed in the previous section with the results obtained using the {\tt Herwig++}~\cite{Herwig} and {\tt Pythia8}~\cite{Pythia8} event generators~\footnote{To be specific, we use {\tt Herwig++ 2.7.0} and {\tt Pythia 8.201} (tune {\tt 4C}) for all our calculations.}, where the quark- and gluon-initiated jets are simulated using the $Z+q$ and $Z+g$ processes at leading order in QCD (with the $Z$ boson subsequently decayed to $\nu \bar{\nu}$). The event samples were generated for proton-proton collisions at the 13 TeV LHC, using the {\tt CTEQ6L1}~\cite{Cteq} parton distribution functions (PDF) for the {\tt Pythia} generators and the default {\tt MRST LO$^{**}$}~\cite{MRST} PDF and UE model for {\tt Herwig++}. Subsequently, we used a modified version of {\tt DELPHES2}~\cite{Delphes} for including detector effects. For observables based on charged tracks to be discussed in the following, we use a minimum $p_T$ threshold of $1$ GeV for each track. All jets are reconstructed with an anti-$k_t$ algorithm~\cite{Cacciari:2008gp,Fastjet} with radius parameter $R=0.4$, and are required to have $p_T > 20$ GeV. In addition, the leading jet is required to be central with $|\eta|<2$. 

\begin{figure}[htb!]
\centering 
\includegraphics[width=0.47\textwidth]{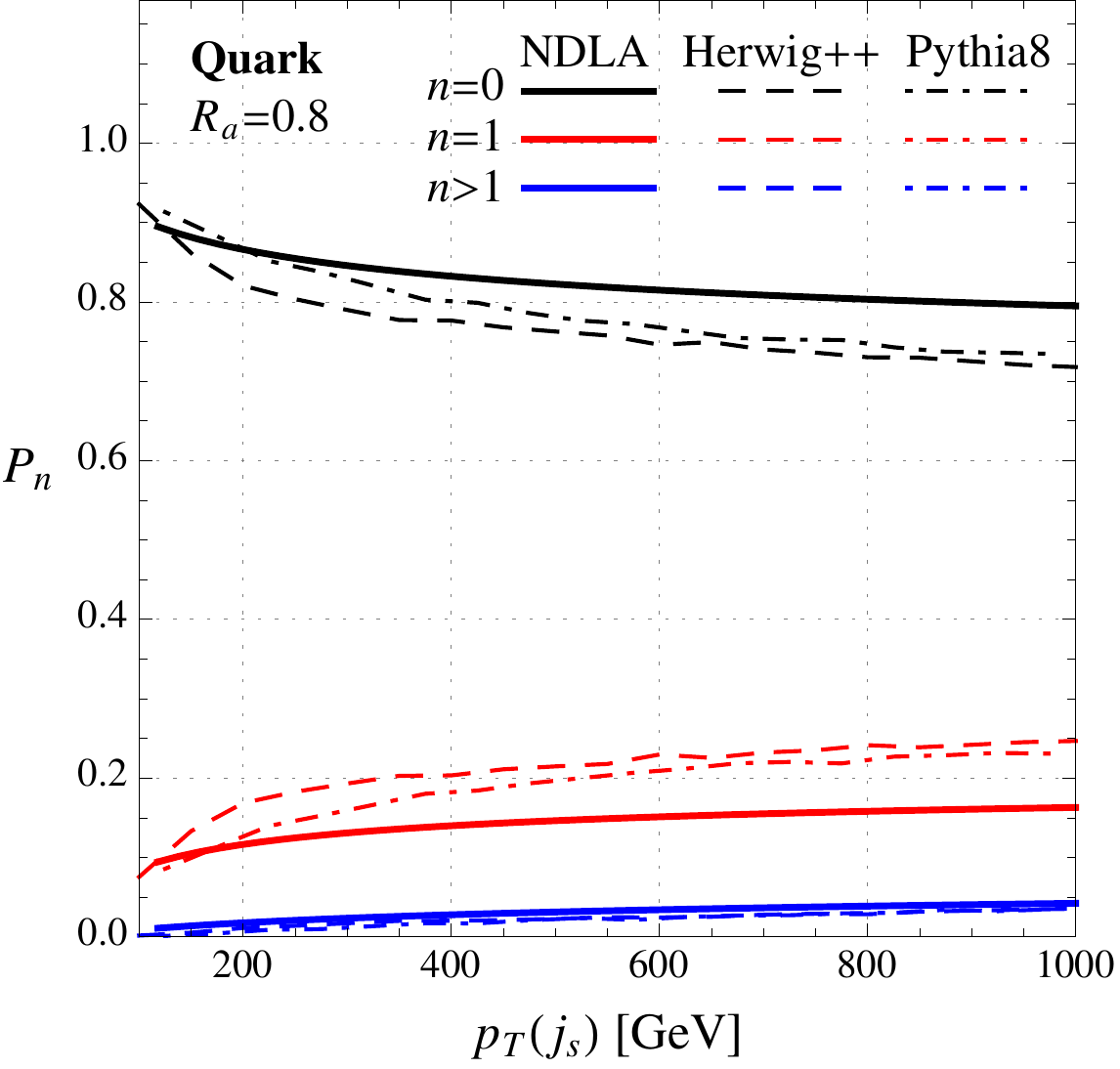}
\hfill
\includegraphics[width=0.47\textwidth]{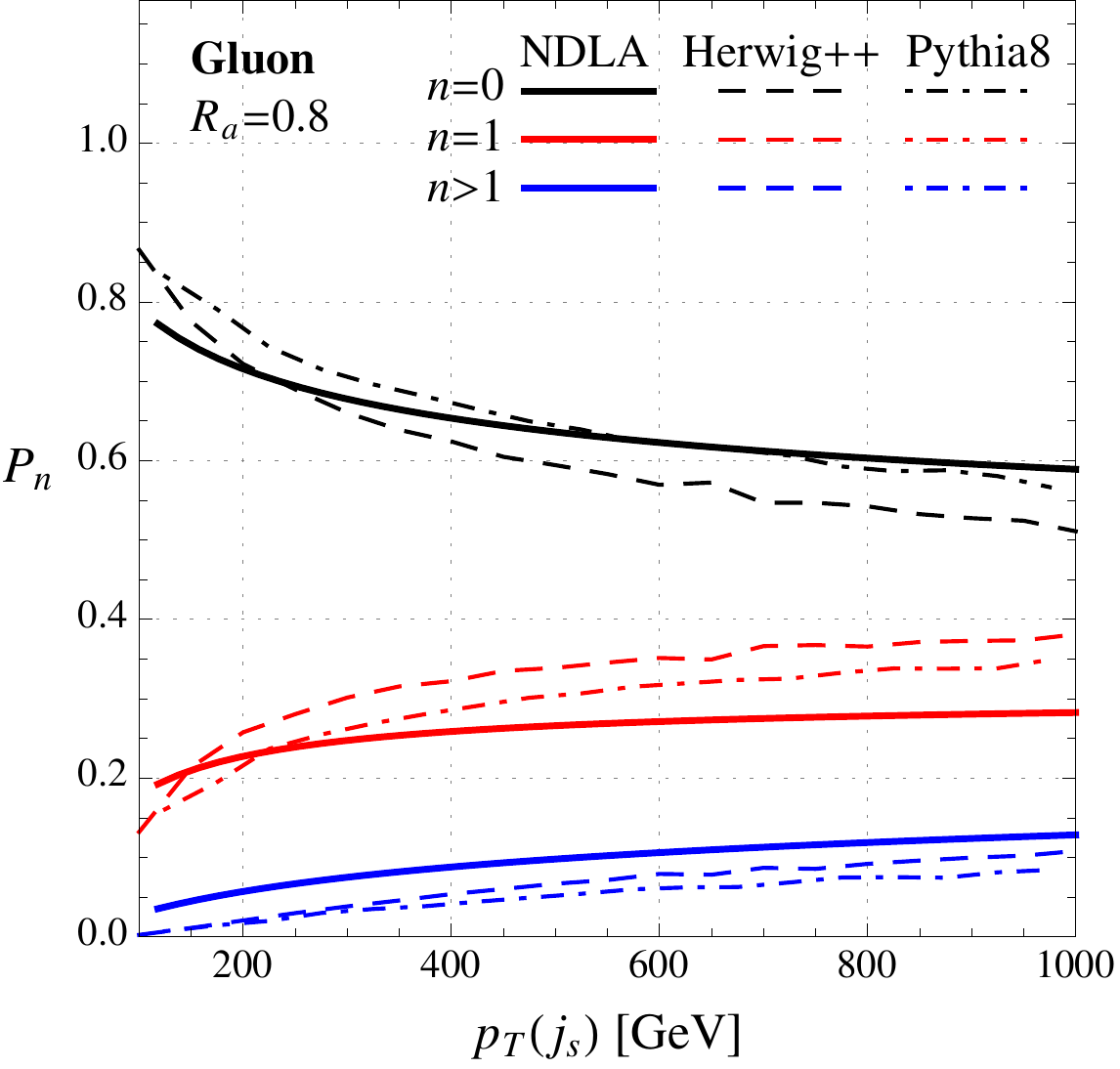}
\caption{\label{fig:1}Comparison of the {\tt Herwig++} and {\tt Pythia8} MC predictions for associated jet rates with the NDLA results, as a function of $p_T(j_s)$: for quark jets (left), and gluon jets (right), with $R_a=0.8$ and $p_a=20$ GeV. Here, $p_T(j_s)$ is the vector sum of the leading jet and associated jet $p_T$'s.}
\end{figure}

In Fig.~\ref{fig:1} we show the probability of obtaining $n$ associated jets $P_n$ as a function of the jet $p_T$ for $n=0,1$ and $n>1$, for quark- and gluon-initiated jets, in the left and right columns respectively. The association radius is set to be $R_a=0.8$ and the minimum associated jet transverse momentum is $p_a=20$ GeV. In the MC simulations, $P_n$ has been computed as a function of $p_T(j_s)$, which is the vector sum of the leading jet and associated jet $p_T$'s. The jet rates are studied as a function of $p_T(j_s)$, as it is closer to the transverse momentum of the parton that initiates the final state shower. 

We see that the functional behaviour with respect to the jet $p_T$ in the MC computation~\footnote{For the associated jet rate calculations, we generated MC event samples with a statistics of 20,000 events each fixing the threshold for the minimum leading jet $p_T$ at $50\times (i+1)$ GeV, for $i \in [0,19]$. Only events with the leading jet $p_T(j_s)$ above the generation threshold are used in the analysis. This ensures uniform MC statistics in the whole range of $p_T(j_s)$.}  and the NDLA calculation are similar, although there are some differences in the values of $P_n$. In particular, the MC prediction of $P_1$ for quark and gluon jets is higher than the NDLA result, especially at higher  $p_T(j_s)$,  with  {\tt Herwig++} giving rise to a slightly larger  $P_1$ compared to {\tt Pythia8}. For a quark jet, the probability of having at least one associated jet ranges from around $15\%$ to $25\%$ as we go from $p_T(j_s)=200$ GeV to $p_T(j_s)=500$ GeV and at higher $p_T(j_s)$ the probability essentially remains the same. For gluon jets, the corresponding probability ranges from around $30\%$ to $40\%$ as we go from $p_T(j_s)=200$ GeV to $p_T(j_s)=500$ GeV. The larger probability to have an associated jet around a gluon can thus be utilized to better discriminate it from quarks, as we shall see in the next section.

\begin{figure}[htb!]
\centering 
\includegraphics[keepaspectratio=true, scale = 0.55]{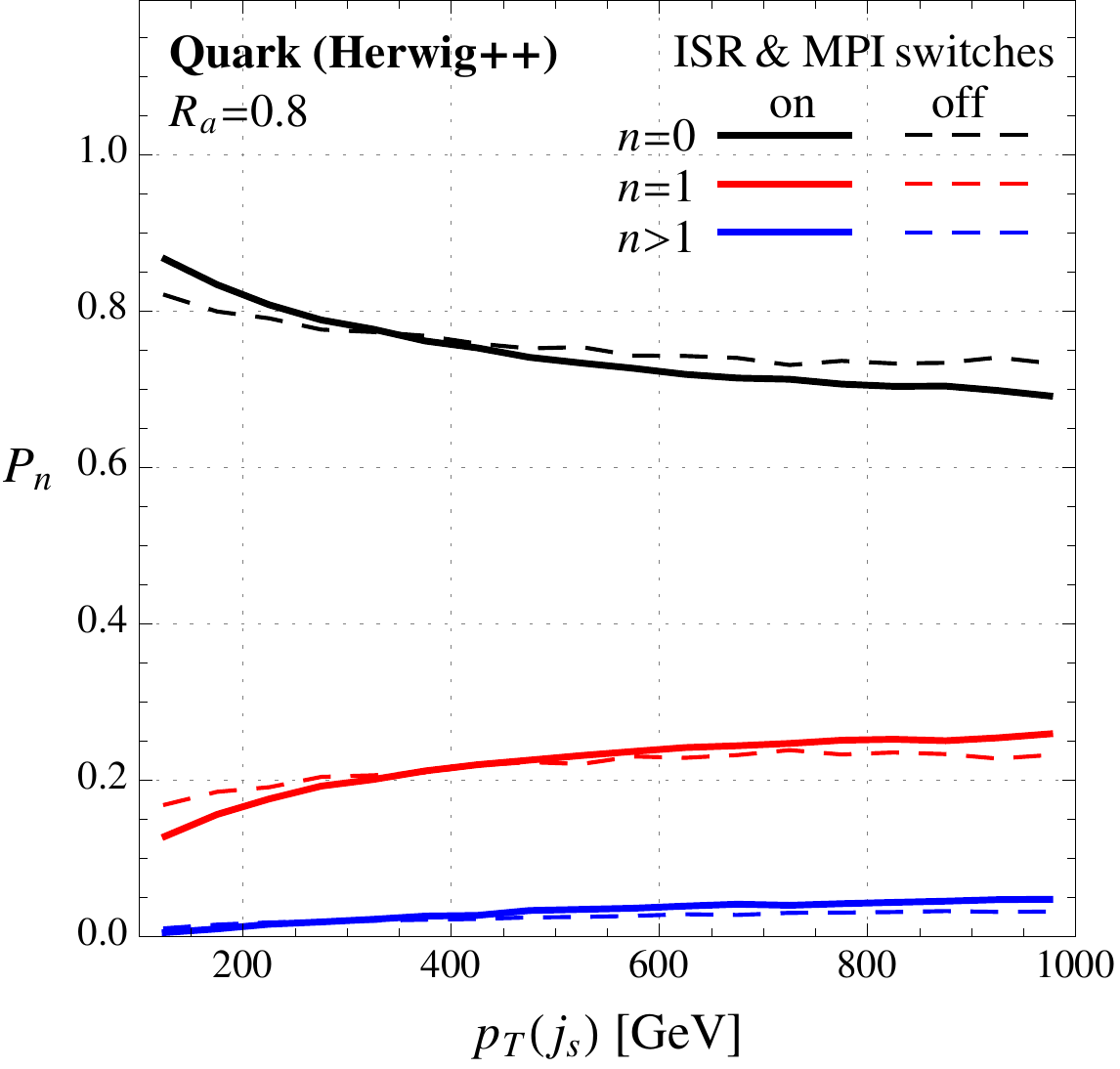}
\hfill
\includegraphics[keepaspectratio=true, scale = 0.55]{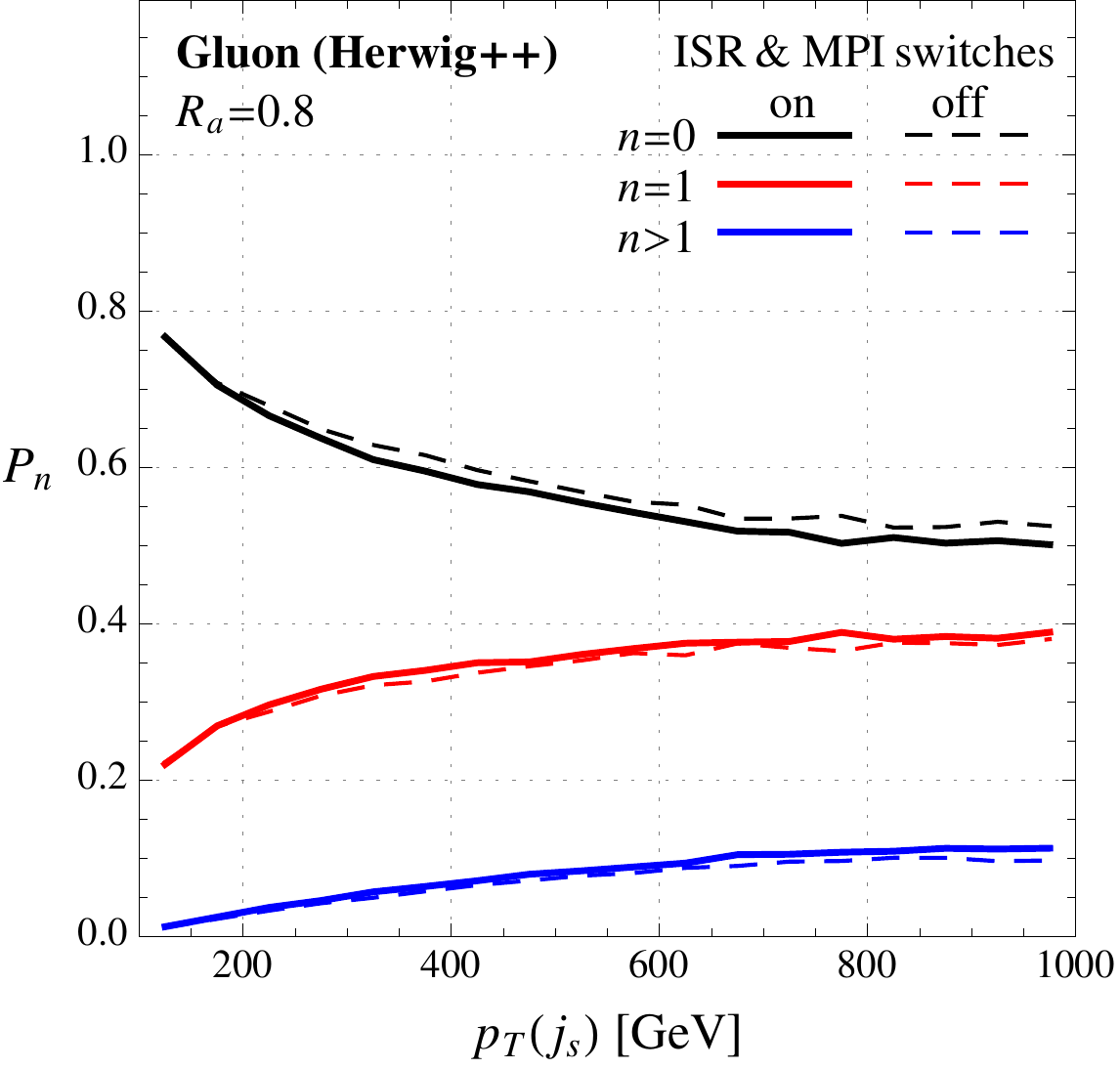}
\includegraphics[keepaspectratio=true, scale = 0.55]{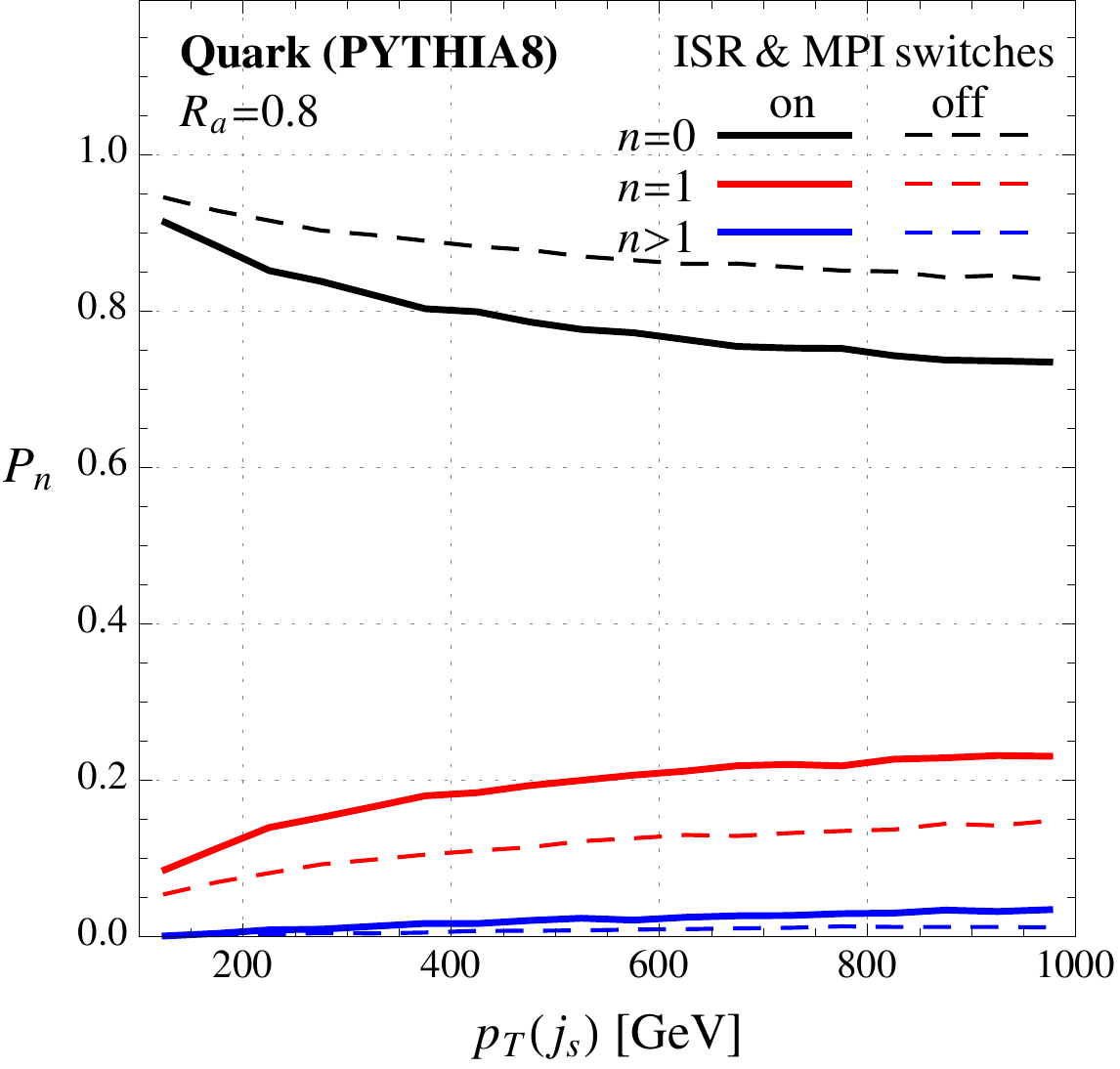}
\hfill
\includegraphics[keepaspectratio=true, scale = 0.55]{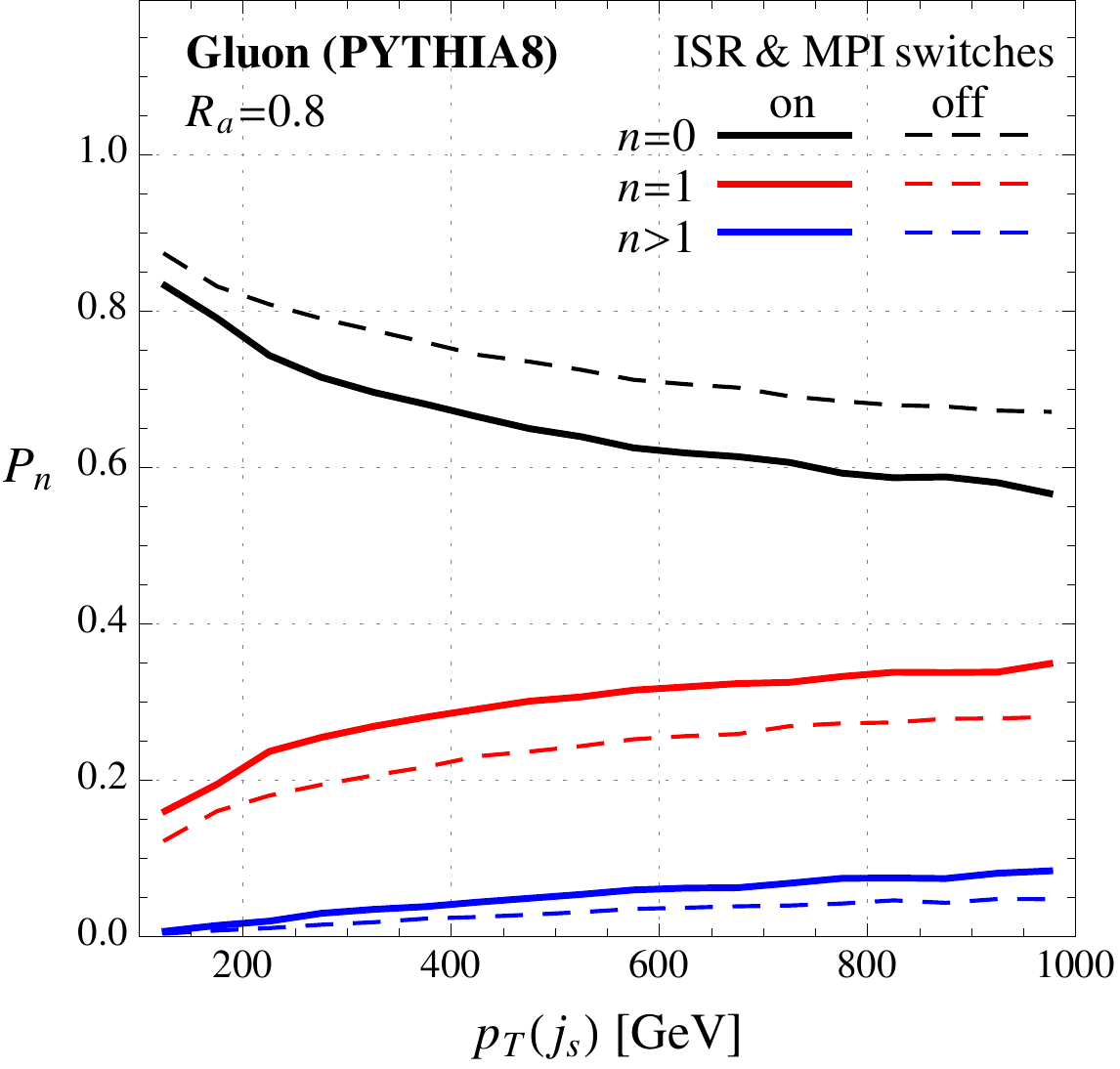}
\includegraphics[keepaspectratio=true, scale = 0.55]{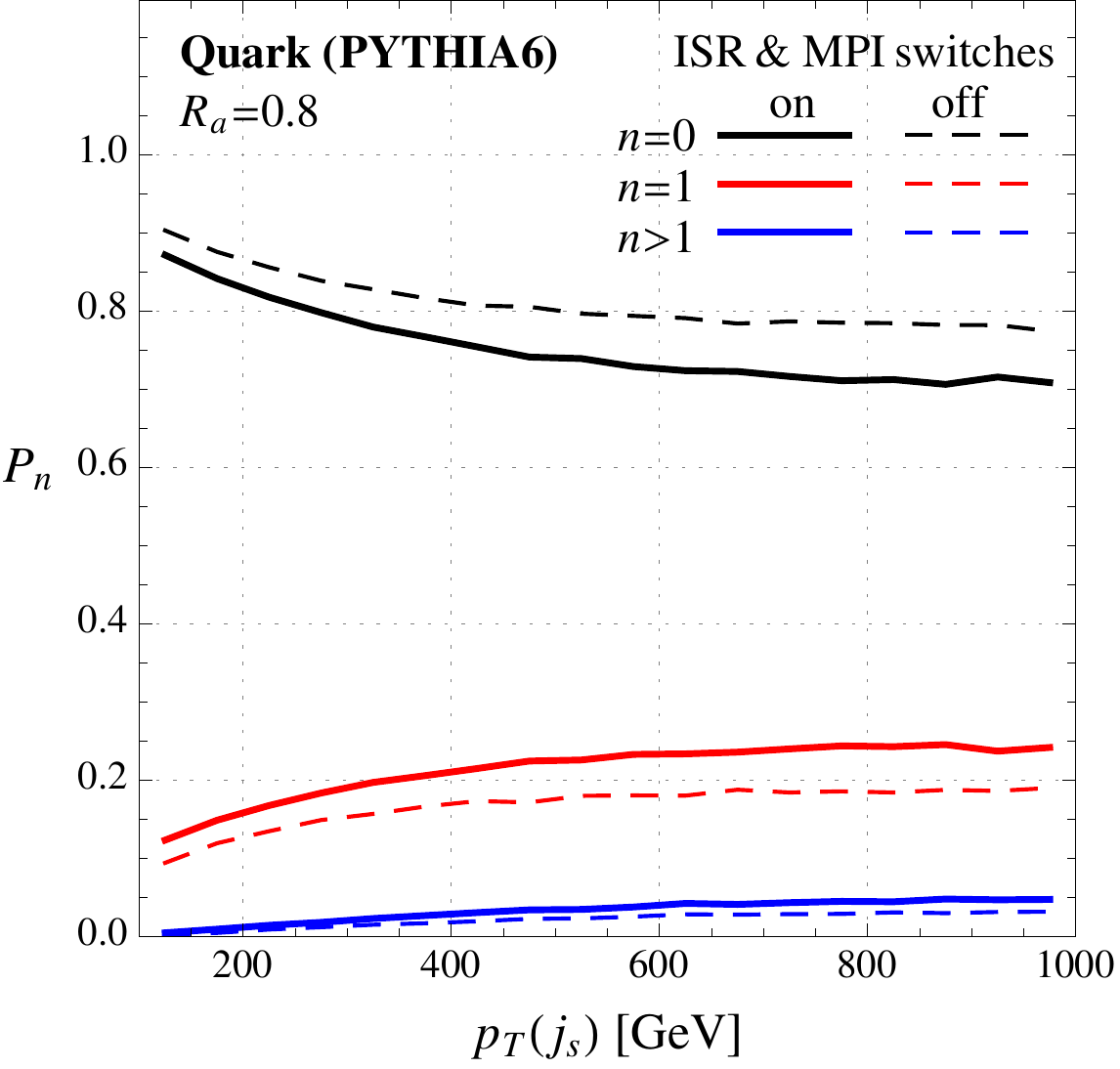}
\hfill
\includegraphics[keepaspectratio=true, scale = 0.55]{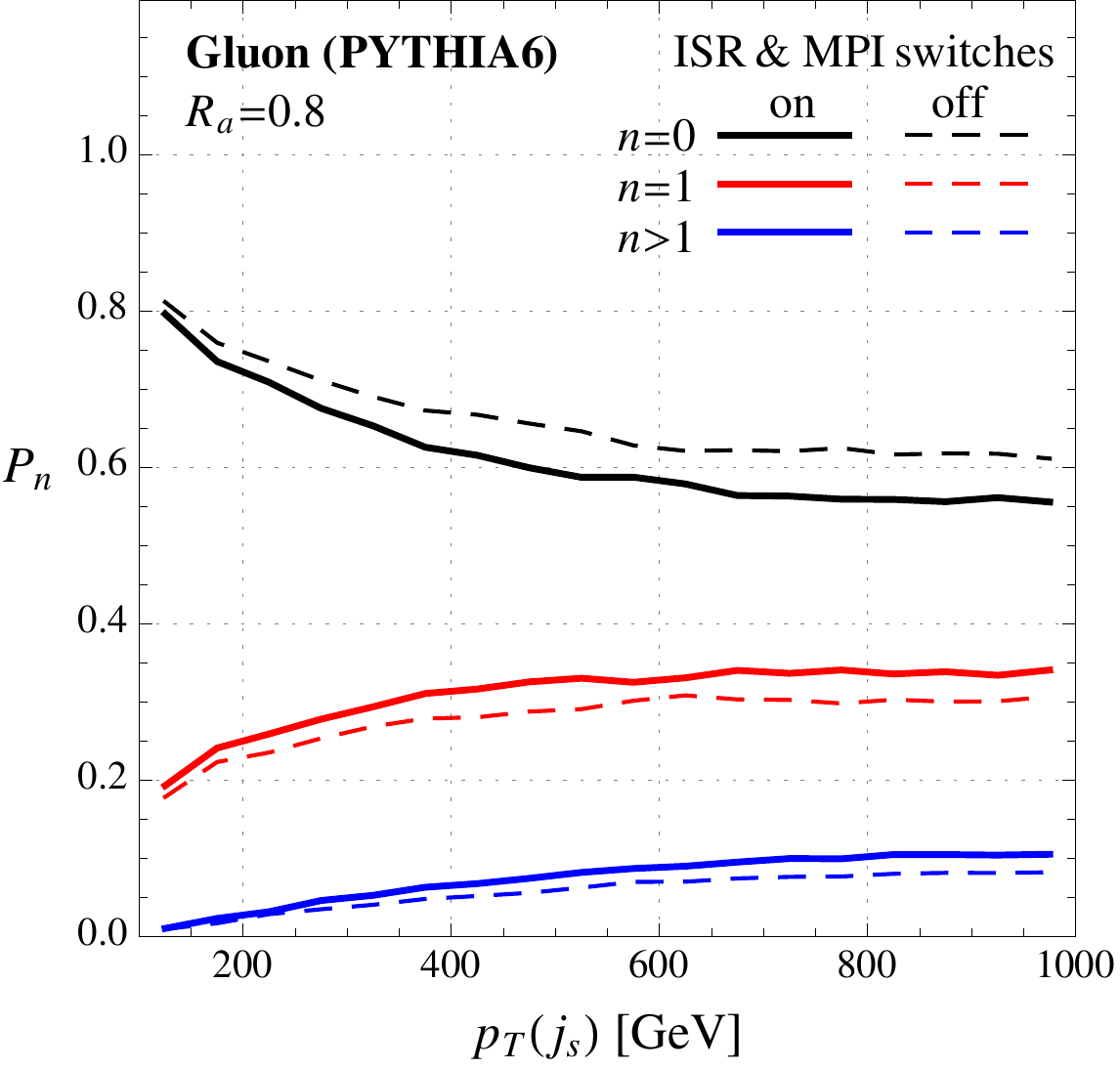}
\caption{\label{fig:2}Comparison of the {\tt Herwig++}, {\tt Pythia8} and {\tt Pythia6} predictions for associated jet rates with and without ISR and MPI, as a function of $p_T(j_s)$: for quark jets (left), and gluon jets (right). Here, $p_T(j_s)$ is the vector sum of the leading jet and associated jet $p_T$'s.}
\end{figure}

The NDLA computation includes only the time-like showering of the final state partons, and ignores some power-suppressed effects due to momentum conservation and hadronization.  On the other hand, the MC results shown above include momentum conservation and hadronization as well as the effects of initial state radiation (ISR) and multiple interaction (MPI). In order to quantify the effect of ISR and MPI, we compare the predictions for $P_n$ with and without ISR and MPI in {\tt Herwig++}, {\tt Pythia8} as well as in {\tt Pythia6}~\cite{Pythia6} (we use the version {\tt Pythia 6.4.28} with the {\tt AUET2B-CT6L} tune) in Fig.~\ref{fig:2}. It is clear from this figure that the impact of ISR and MPI is rather small for our choice of the association radius $R_a=0.8$, thereby making the predictions stable against such effects. For this choice of $R_a$, we can see that {\tt Pythia8} shows the highest variation against such effects, followed by {\tt Pythia6}, while the effects are indeed negligible for the case of {\tt Herwig++}~\footnote{However, we have checked that if we take a larger association radius, $R_a > 1.2$, the ISR effects become appreciable in {\tt Herwig++}.}. Furthermore, the MC results become closer to the NDLA ones when ISR and MPI effects are switched off.

We also investigated the effects of momentum conservation, by changing the recombination scheme in the anti-$k_t$ jet algorithm from the default $E$-scheme
to the ``winner-take-all'' scheme introduced in~\cite{Bertolini:2013iqa}, which
is less sensitive to recoils in the parton shower~\cite{Larkoski:2014uqa}.
Such a change increases the MC associated jet rates very slightly. We believe this is because the axis of the leading jet is moved away from the overall momentum vector of the system. The effects are roughly proportional for quark and gluon jets, so they would not affect discrimination significantly. 

\section{Quark-gluon separation: multivariate analysis }
\label{sec:4}
\subsection{Variables for quark-gluon separation}
A large number of variables have been surveyed in the context of quark-gluon discrimination, constructed out of either track based observables or calorimeter based ones~\cite{Schwartz1,Schwartz2,Schwartz3,Larkoski1,Larkoski2}. While the former category has the practical advantage of being more accurate due to better track momentum resolution as well as being less prone to pile-up contamination, the latter category can be used for jets with larger rapidities outside the tracker coverage. The most widely studied variables include the number of charged tracks inside the jet cone ($n_{\rm ch}$), the jet width~\cite{Schwartz1} and energy-energy-correlation (EEC) angularity~\cite{Larkoski1}. The jet width is defined as
\begin{equation}
w = \frac{\sum_{i} p_{T,i} \times \Delta R(i,{~\rm jet})}{\sum_{i} p_{T,i} }
\end{equation}
where the sum goes over all the tracks associated to the jet. A similar track-based EEC variable, denoted by $C_1^{(\beta)}$ can be defined as
\begin{equation}
C_1^{(\beta)}=\frac{\sum_{i} \sum_{j} p_{T,i} \times p_{T,j} \times (\Delta R(i,j))^\beta}{(\sum_{i} p_{T,i} )^2}.
\end{equation} 
Here again the sum over $i$ and $j$ run over all the tracks associated to the jet with $j>i$, while $\beta$ is a tunable parameter. It has been demonstrated in Ref.~\cite{Schwartz3,Larkoski1} that smaller values of the exponent $\beta$ leads to a better quark-gluon separation, and $\beta=0.2$ is found to be optimal from perturbative calculations and MC studies based on {\tt Herwig++} and {\tt Pythia8} generators. We have compared the performance of the jet width variable $w$ and the EEC variable $C_1^{(\beta=0.2)}$ in the multivariate analyses (MVA) to be discussed below, and find that in all cases $C_1^{(\beta=0.2)}$ leads to a better separation of gluons from quarks. Therefore, in the following, we only show results based on $n_{\rm ch}$ (with each charged track having $p_T>1$ GeV) and $C_1^{(\beta=0.2)}$. In addition, we shall include the associated jet information as well as the jet mass variable and compare the performance of the different MVA methods. As seen in the previous section, for $n=1$ or $n>1$, the probability of finding $n$ associated jets, $P_n$, is significantly larger for gluon jets compared to quark-initiated ones across the whole $p_T$ range of interest. Therefore, the presence (or absence) of an associated jet within a certain distance $R_a$ of a high-$p_T$ jet can be used to further improve the separation.

As the boundary between the signal and background regions in the hyper-surface spanned by the variables is non-linear, it is beneficial to adopt a multivariate analysis strategy as compared to a cut-based one. For this purpose, we employed a Boosted Decision Tree (BDT) algorithm with the help of the {\tt TMVA-Toolkit}~\cite{TMVA} in the {\tt ROOT} framework. The training of the classifier was performed with $Z+q-$jet and $Z+g-$jet samples, and we generated the above MC samples uniformly distributed in jet-$p_T$~\footnote{The MC event samples for the training of the classifier were generated in the same manner as for the associated jet rate computation in the previous section, but with a smaller step size of $10$ GeV for the minimum $p_T(j_s)$ thresholds.}. The input variables for the two variable training are taken to be $n_{\rm ch}$ and $C_1^{(\beta=0.2)}$, while for three-variable trainings we further include the variable $m_J/p_{T,J}$, where $m_J$ is the jet mass and $p_{T,J}$ is the transverse momentum of the leading jet. The information on the number of associated jets is included in the form of two categories ($n=0$ or $n \geq 1$) in the MVA. 

It should be emphasized that the MC prediction of the discrimination variables, especially the number of charged tracks $n_{\rm ch}$ is quite sensitive not only to the parton shower (PS) algorithm adopted and the related parameters, but also to the tuning of the hadronization and underlying event models. This is expected, since $n_{\rm ch}$ is not an infrared safe quantity, and only the ratio $n_{\rm ch}^{\rm gluon}/n_{\rm ch}^{\rm quark}$ converges rather slowly to the ratio of the colour factors $C_A/C_F$ for high jet $p_T$~\cite{Bolzoni}. The disagreement between different MC's can therefore be reduced only by appropriate tuning at the LHC energies. With this limitation of the MC predictions in view, in this study, we compare the performance of different MVA methods within the same MC generator to estimate the improvement in adding associated jet related observables. We also show the quark-gluon separation as predicted by the different MC's for comparison.  In Appendix~\ref{appx} we present details of the distributions of the discrimination variables and the differences
between the MC predictions for them.

\subsection{Performance in MVA}
\begin{figure}[htb!]
\centering 
\includegraphics[width=0.45\textwidth]{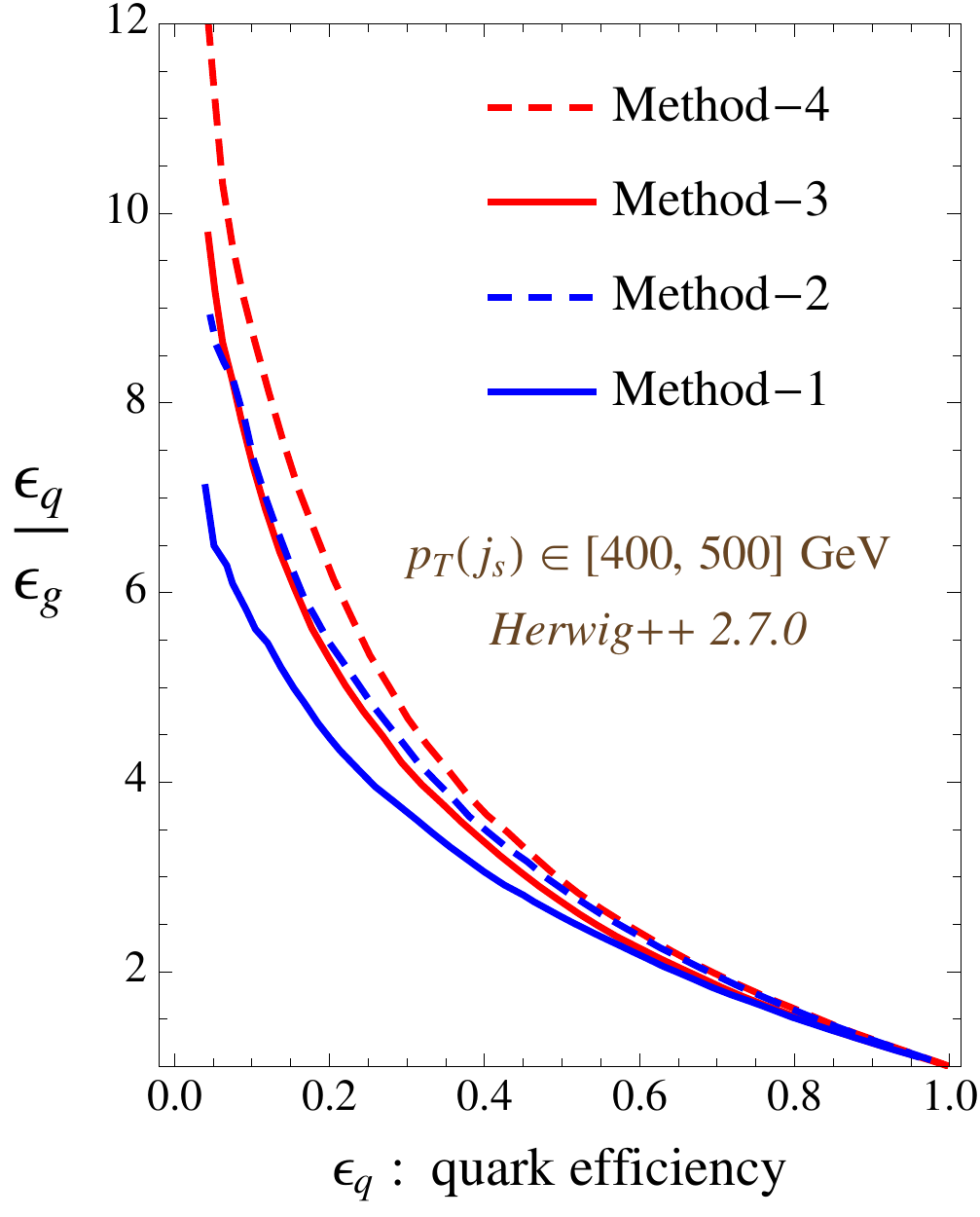}
\hfill
\includegraphics[width=0.45\textwidth]{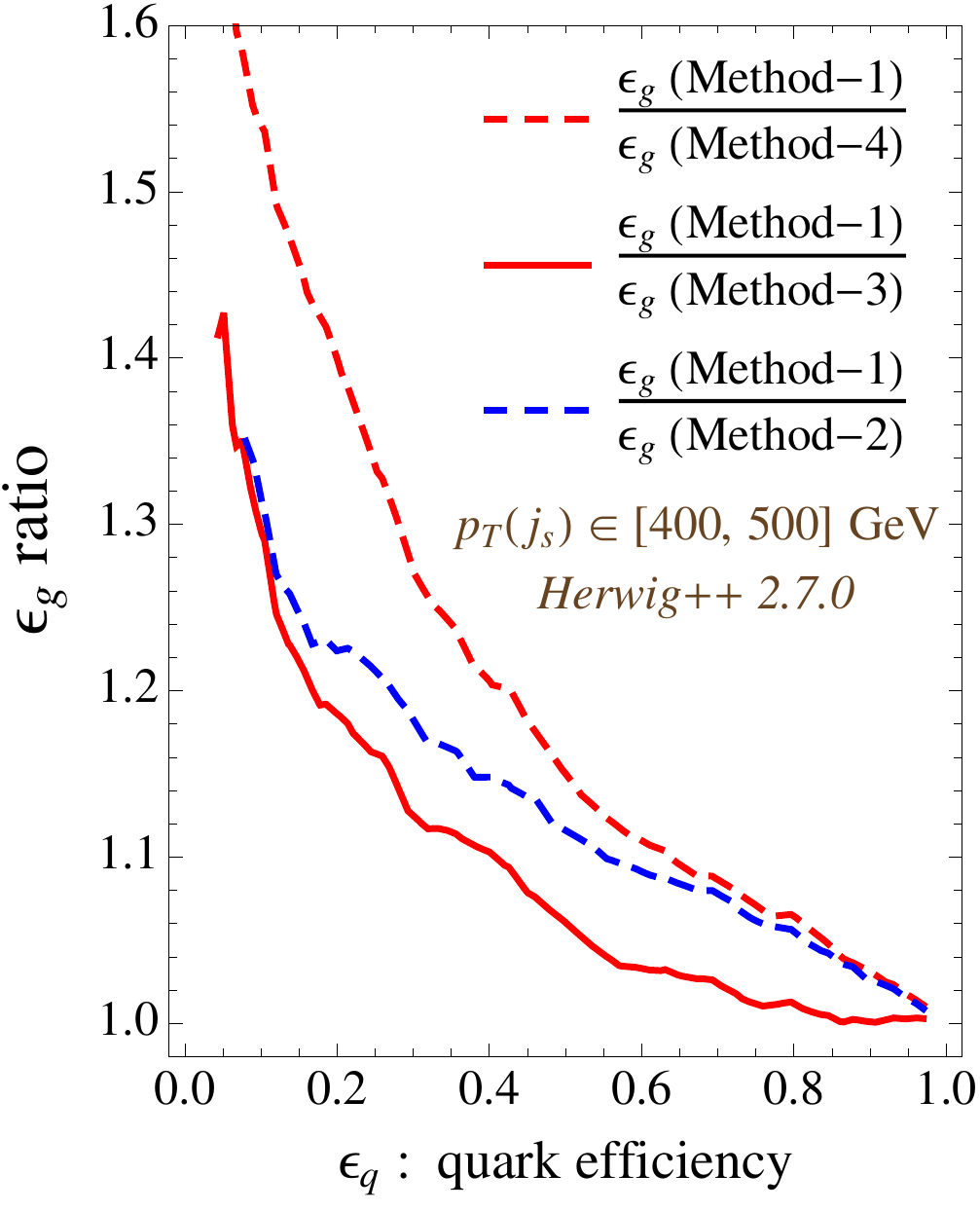}
\caption{\label{fig:3} The ratio of the quark and gluon tagging efficiencies, $\epsilon_q/\epsilon_g$ as a function of $\epsilon_q$, for $400<p_T(j_s)<500$ GeV, as determined by MC simulations with {\tt Herwig++} (left column). The different MVA methods, determined in terms of the input variables are explained in the text. To quantify the improvement in quark gluon separation as we go to Methods 2,3 and 4, we show $\epsilon_g$(Method-1)$/\epsilon_g$(Method-\{2,3,4\})  as a function of $\epsilon_q$ as well (right column).}
\end{figure}

\begin{figure}[htb!]
\centering 
\includegraphics[width=0.45\textwidth]{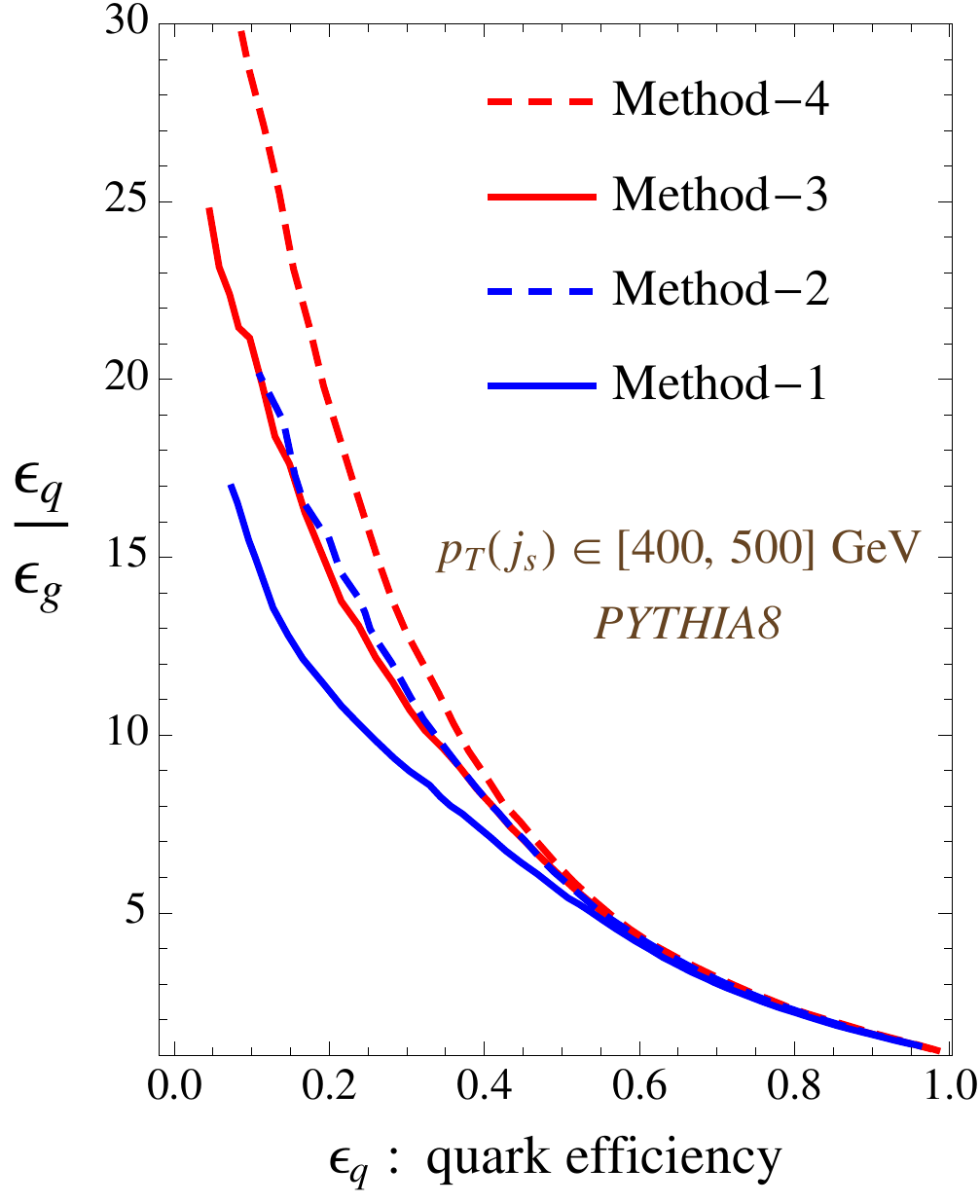}
\hfill
\includegraphics[width=0.45\textwidth]{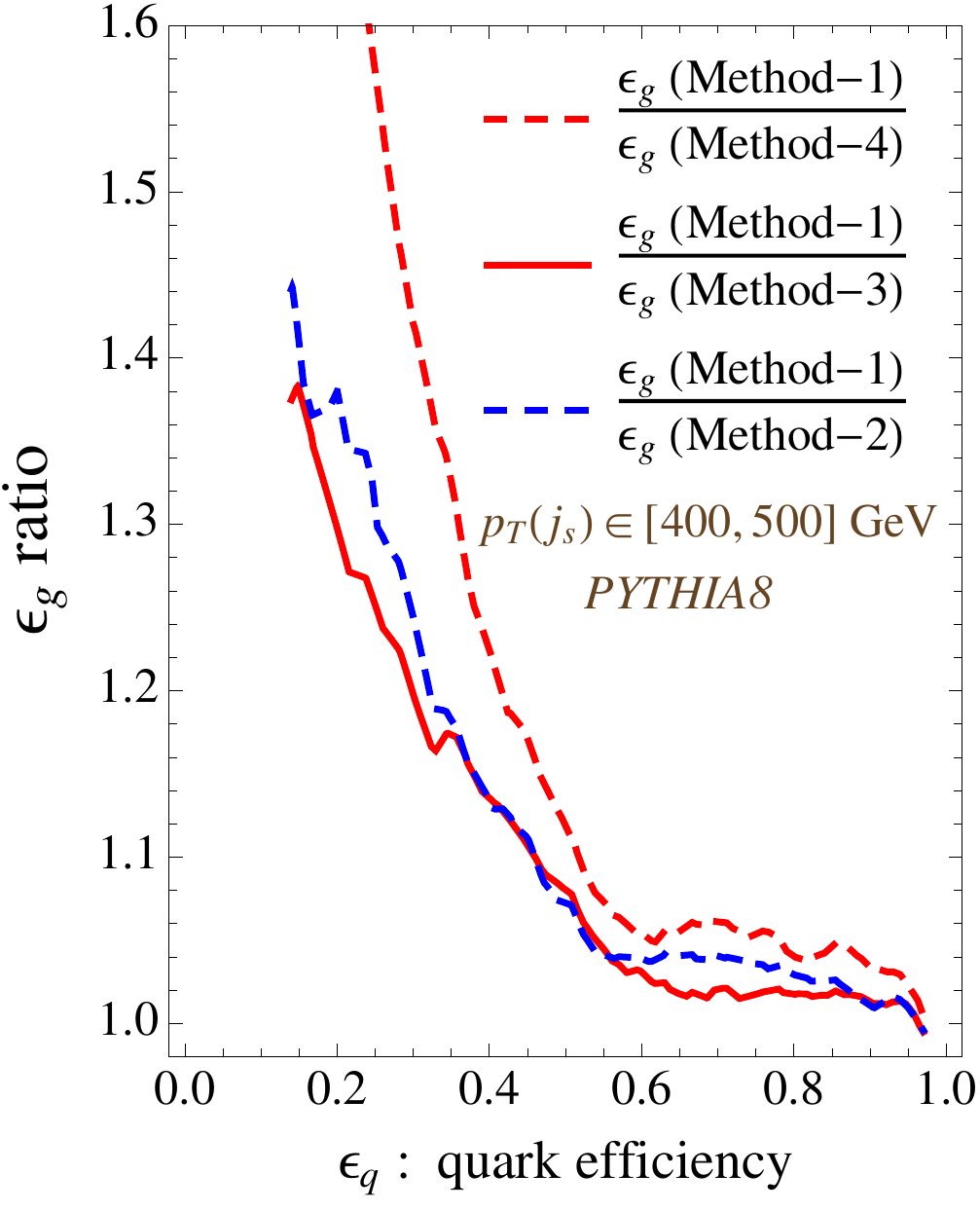}
\caption{\label{fig:3a} Same as Fig.~\ref{fig:3}, with MC simulations using {\tt Pythia8}.}
\end{figure}

\begin{figure}[htb!]
\centering 
\includegraphics[width=0.45\textwidth]{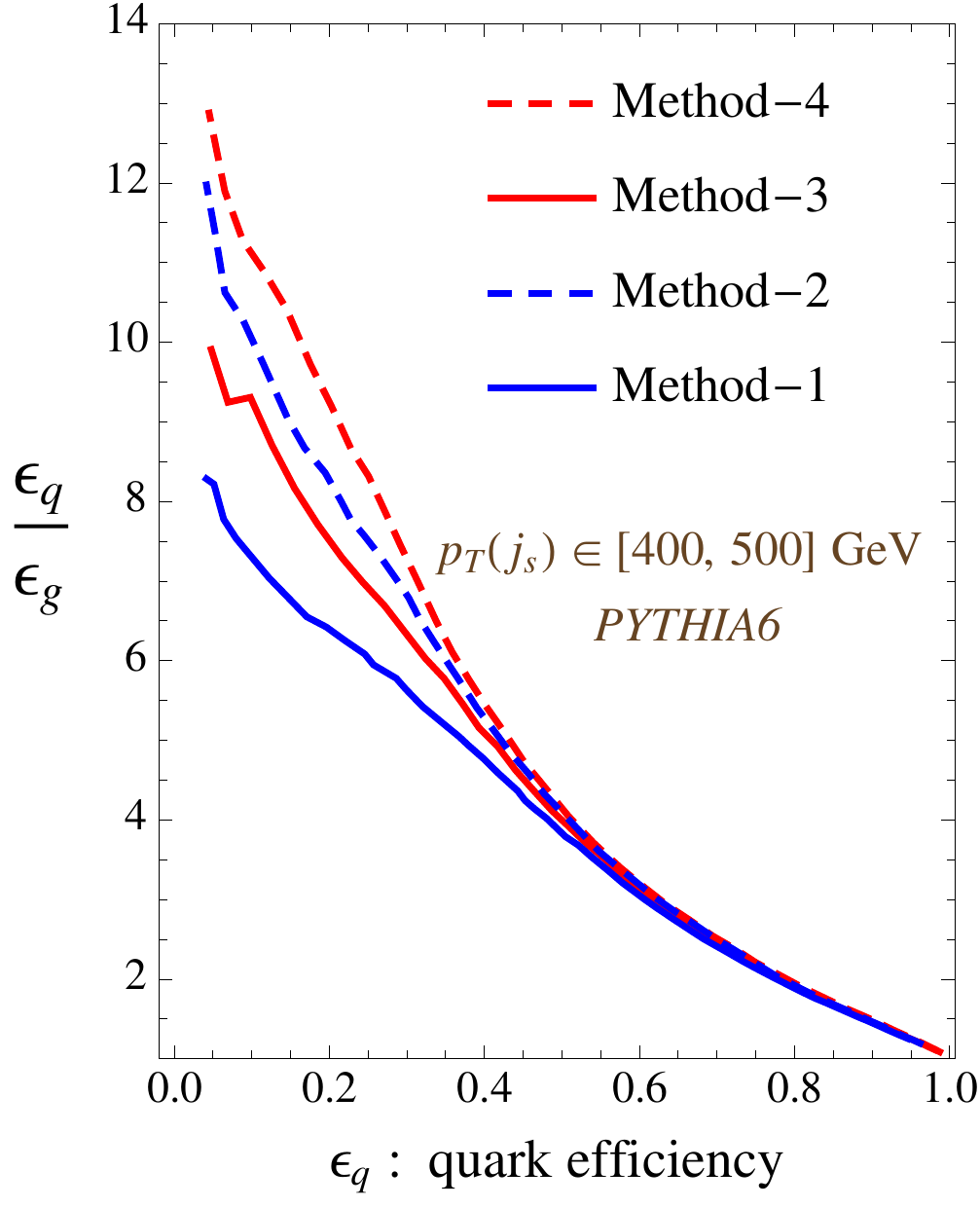}
\hfill
\includegraphics[width=0.45\textwidth]{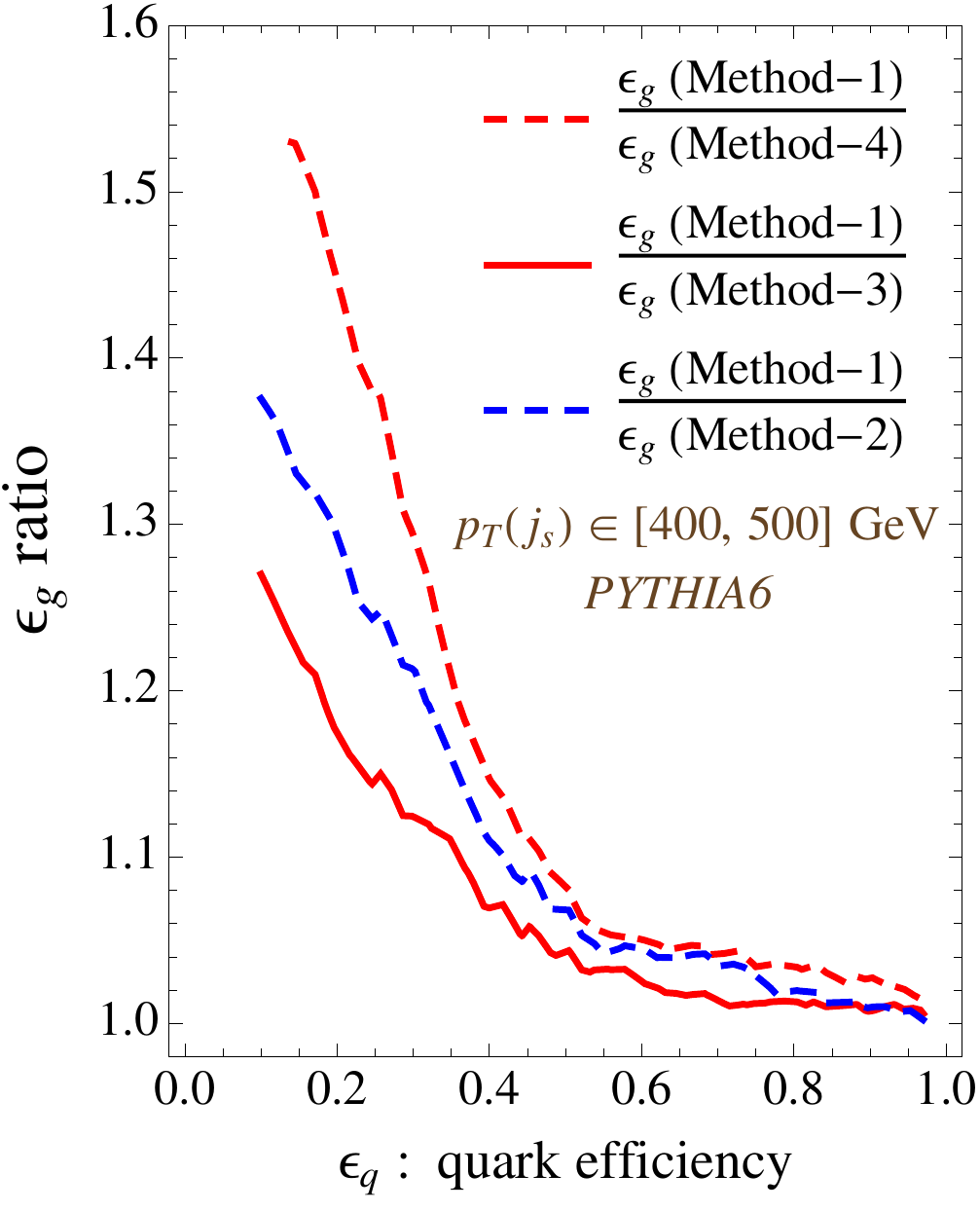}
\caption{\label{fig:3b} Same as Fig.~\ref{fig:3}, with MC simulations using {\tt Pythia6}.}
\end{figure}

Based on the BDT analysis, we obtain the efficiencies of tagging quark ($\epsilon_q$) and gluon jets ($\epsilon_g$) as a function of the cut on the BDT score. It is more useful to compare the ratio of the tagging efficiencies as a function of $\epsilon_q$, in order to judge the separation power of a "quark-rich signal" from a "gluon-rich" background. In Figs.~\ref{fig:3}-\ref{fig:3b} (left column) we show the ratio of the quark and gluon tagging efficiencies, $\epsilon_q/\epsilon_g$ as a function of $\epsilon_q$, for $400<p_T(j_s)<500$ GeV, with the event samples generated with all the three MC codes. Four different MVA methods are shown corresponding to different choices for the discrimination variables:
\begin{itemize}
\item {\bf Method-1:} Two variables, $n_{\rm ch}$ and $C_1$ with $\beta=0.2$.
\item {\bf Method-2:} Two variables, $n_{\rm ch}$ and $C_1$ with $\beta=0.2$,  with two categories determined in terms the number of associated jets ($n=0$ or $n \geq 1$).
\item {\bf Method-3:} Three variables, $n_{\rm ch}$, $C_1$ with $\beta=0.2$ and $m_J/p_{T,J}$.
\item {\bf Method-4:} Three variables, $n_{\rm ch}$, $C_1$ with $\beta=0.2$ and $m_J/p_{T,J}$, with two categories determined in terms the number of associated jets ($n=0$ or $n \geq 1$).
\end{itemize}

We can quantify the improvement in quark-gluon separation using $\epsilon_g$(Method-1)$/\epsilon_g$(Method-\{2,3,4\})  as a function of $\epsilon_q$, as shown in Figs.~\ref{fig:3}-\ref{fig:3b} (right). For e.g., for an operating point of $\epsilon_q=0.4$, we can obtain an improvement of around $10\%, 15\%$ and $20\%$ using Methods-2,3 and 4 respectively, when compared to Method-1. The differences between the improvement factors obtained using the three MC's are found to be small.

\begin{figure}[htb!]
\centering 
\includegraphics[width=0.45\textwidth]{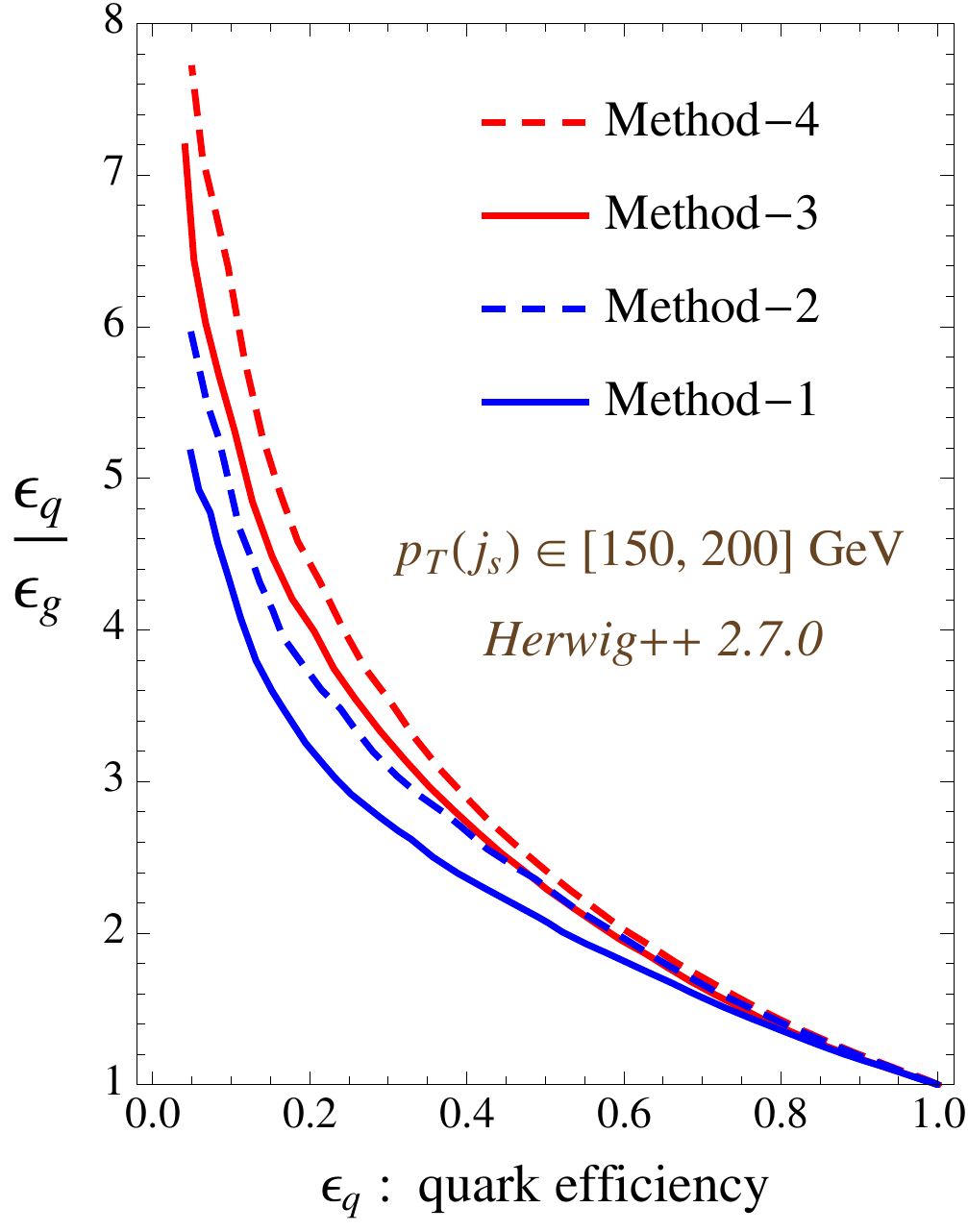}
\hfill
\includegraphics[width=0.45\textwidth]{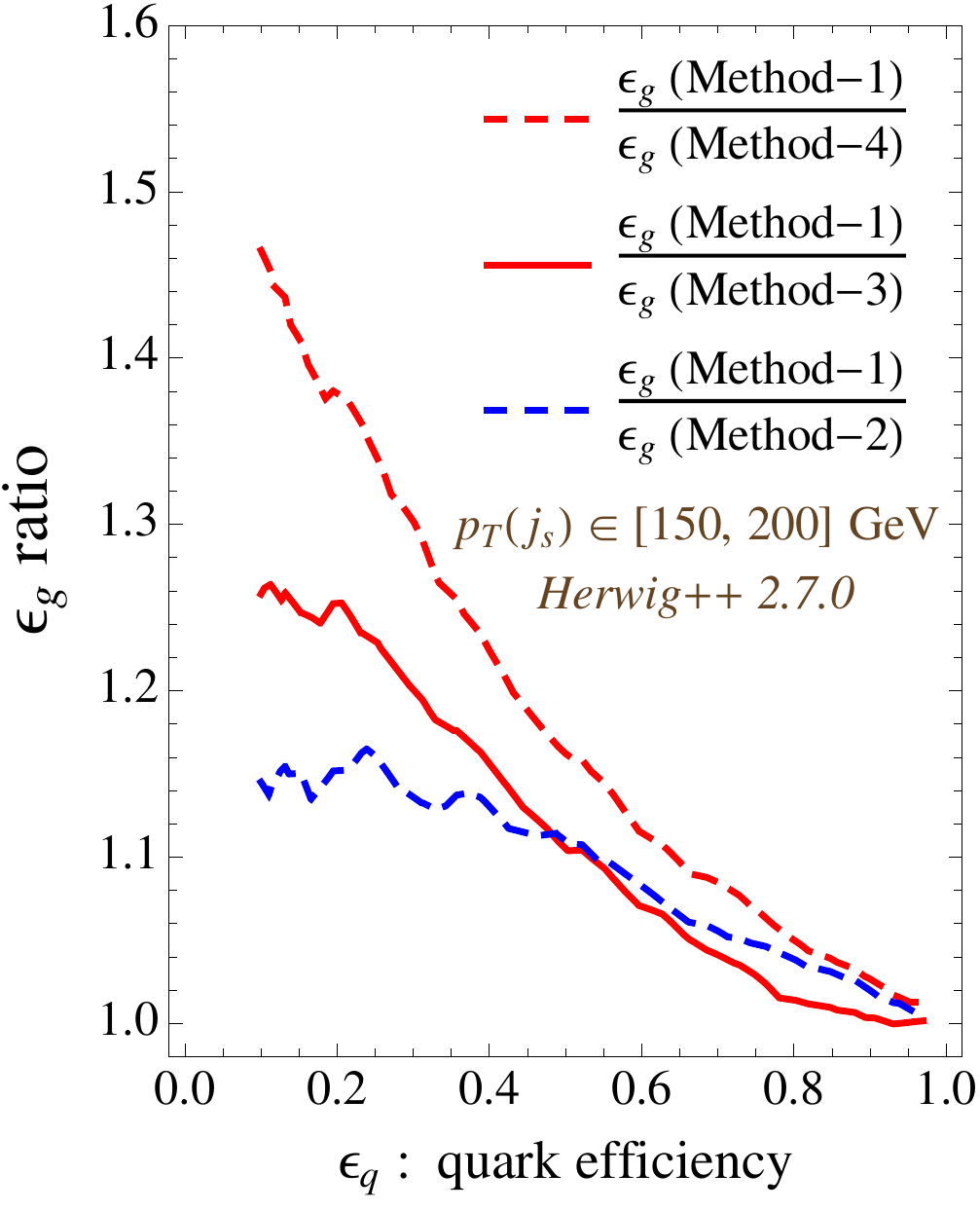}
\caption{\label{fig:4} Same as Fig.~\ref{fig:3}, for a lower range of jet $p_T$, $150<p_T(j_s)<200$ GeV. Results using only {\tt Herwig++} are shown.}
\end{figure}

In order to estimate the change in tagger performance as we consider lower $p_T$ jets, we show in Fig.~\ref{fig:4} the same results as in Fig.~\ref{fig:3}, but now with $150<p_T(j_s)<200$ GeV. The improvement on adding associated jet rates is still appreciable, although it is somewhat reduced compared to the higher $p_T$ range. The fluctuations in the $\epsilon_g$ ratio for lower values of $\epsilon_q$ in Fig.~\ref{fig:4} are due to low MC statistics.

We can see in Figs.~\ref{fig:3}-\ref{fig:3b} that there is an improvement in going from a two variable analysis to a three variable one by including the variable $m_J/p_{T,J}$. This can be understood as follows. The jet mass variable is related to $C_1^{(\beta=2)}$, as can be seen by writing both of them in terms of the $z,\theta$ variables for the hardest emission inside the jet cone: $m_J^2 \simeq z(1-z) \theta^2 p_{TJ}^{2}$. Furthermore, $C_1^{(\beta=2)}$ and $C_1^{(\beta=0.2)}$ are two independent variables belonging to the $C_1$ class which carry all the information on this hardest emission, and including both of them improves the tagger performance. For this reason, further addition of a third variable in the $C_1$ class does not change the performance appreciably, a fact that we explicitly checked by a separate MVA analysis. There is a further improvement in the quark-gluon separation when the number of associated jets information is included at the level of categories in both the two and three variable MVA analyses. Since the associated jet rates carry the additional information of radiation outside the jet cone, Methods 2 and 4 lead to further improvements as compared to Methods 1 and 3, respectively.

\begin{figure}[htb!]
\centering 
\includegraphics[width=0.45\textwidth]{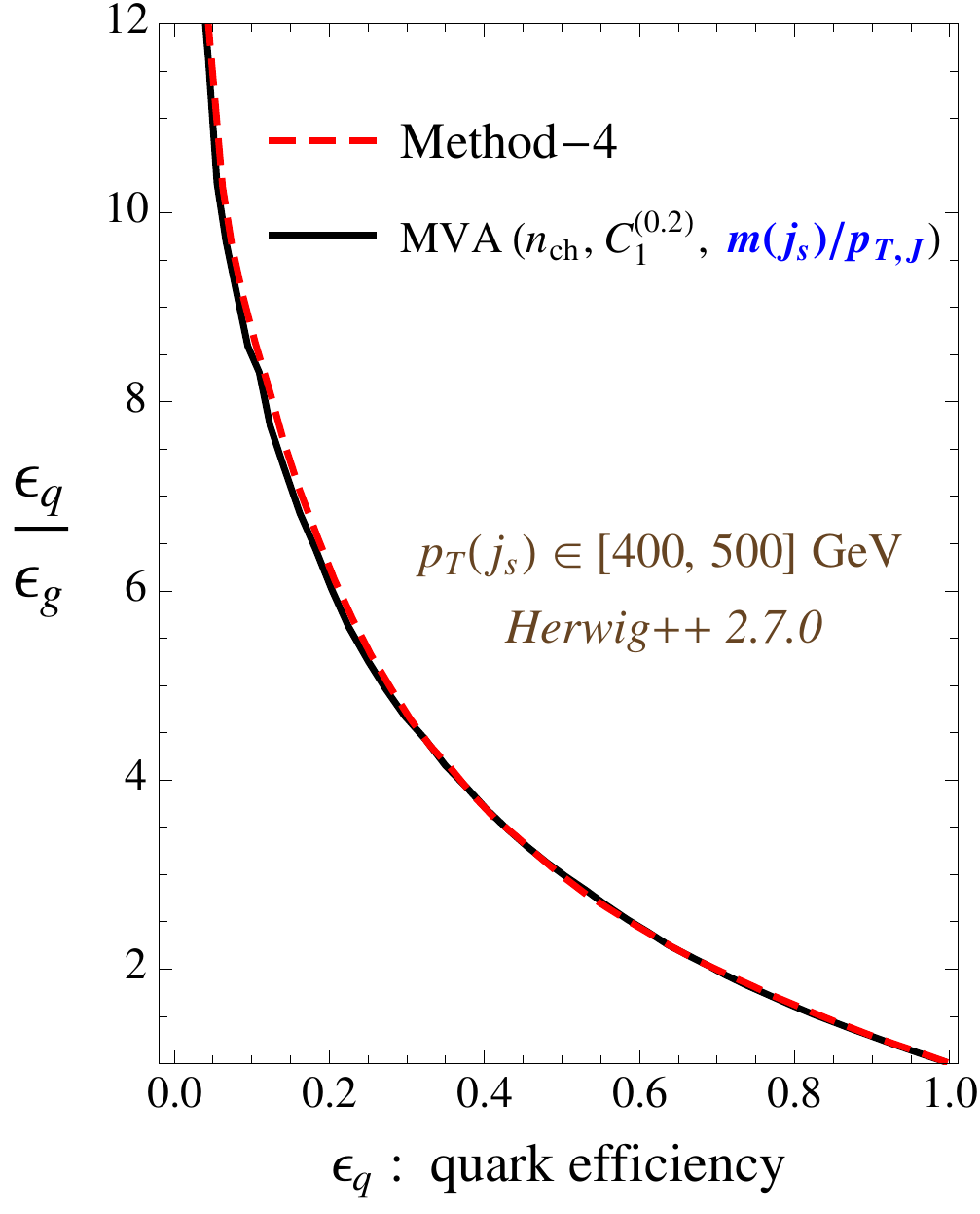}
\caption{\label{mjs} Comparison of Method 4 which includes $m_J/p_{T,J}$ and the associated jet rates as categories in the MVA, and the alternative method of including the associated jet rate information by using the modified jet mass variable $m(j_s)/p_{T,J}$. Both methods lead to the same MVA performance.}
\end{figure}

Method 4 leads to the best performance out of the four different MVA's considered. In fact, we find that there is an alternative way to include the associated jet rates information in Method 4 by using the modified jet mass variable $m(j_s)/p_{T,J}$ in Method 3. Here, $m(j_s)$ is the jet mass computed by adding the leading jet and associated jet four momenta. Because of a larger associated jet rate, for the same $p_T(j_s)$, $m(j_s)$ is higher for a gluon jet compared to a quark, while $p_{T,J}$ is lower. Therefore, using either associated jet rate categories and $m_J/p_{T,J}$, or using only the variable $m(j_s)/p_{T,J}$ leads to the same MVA performance, as shown in Fig.~\ref{mjs}.

\section{Subjet rates in jets: analytical calculations}
\label{sec:5}
The number of charged tracks inside a jet cone, $n_{\rm ch}$ (with each track having transverse momentum above a threshold, usually taken to be around $1$ GeV) is often used as a good discriminating variable. However, as mentioned earlier, the MC predictions for this observable are quite sensitive not only to the parton shower (PS) algorithm and the related parameters, but also to the tuning of the hadronization and underlying event models. On the otherhand, we find that the number of subjets of a primary jet leads to a more uniform prediction across the MC's, and thus can be better suited in quark gluon separation studies.
The number of subjets as a quark-gluon separation variable was considered earlier in Ref.~\cite{Schwartz1}. In this study, we compute the subjet rates to NDLA accuracy, and show a detailed comparison with different MC's.
 
We find the {\it subjets} of jet $j$ with the {\it exclusive $k_t$ algorithm}, which applies the dimensionless distance measure
\beq
y_{ik} = \min\{p_{ti}^2,p_{tk}^2\}\frac{\Delta R_{ik}^2}{R^2 p_{j}^2}\;,
\eeq
to its constituent objects and clusters them as discussed for a  generalized $k_t$ algorithm in Sec.~\ref{sec:2}, until the smallest
$y_{ik}$ is above $y_{\rm cut}$.  Thus the subjet rates are functions
of the jet $p_t=p_j$, the jet radius $R$, and $y_{\rm cut}$.

 In this section, we compute the subjet rates to NDLA, i.e.\ considering
double and next-to-double logarithms, $\as^n L^{2n}$ and
$\as^n L^{2n-1}$, where now $L=\ln(1/y_{\rm cut})$.   The relevant
generating functions in this case are those given in
Refs.~\cite{Catani:1991hj,Ellis:1991qj}:
\beqn
\phi_q(u,Q) &=& u\,\Delta_q(Q) \exp\left(\int_{Q_0}^{Q} dq
  \,\Gamma_q(Q,q)\phi_g(u,q)\right)\,,\\
\phi_g(u,Q) &=& u\,\Delta_g(Q) \exp\left(\int_{Q_0}^{Q} dq
  \left[\Gamma_g(Q,q)\phi_g(u,q)
+\Gamma_f(q)\frac{\phi_q(u,q)^2}{\phi_g(u,q)}\right]\right)
\eeqn
where $Q=R\,p_j$ is the jet scale, $Q_0=R\,p_j\sqrt y_{\rm cut}$ is the
resolution scale,\footnote{Here again we keep power-suppressed
  corrections in order to satisfy boundary conditions.}

\beqn
\Gamma_q(Q,q) &=& \frac{2C_F}{\pi}\frac{\as(q^2)}{q}\left(\ln\frac
  Q{q} -\frac 34+\frac{q}Q -\frac 14\frac{q^2}{Q^2}\right)\,,\\
\Gamma_g(Q,q) &=& \frac{2C_A}{\pi}\frac{\as(q^2)}{q}\left(\ln\frac
  Q{q} -\frac{11}{12}+\frac{q}Q -\frac 14\frac{q^2}{Q^2}
+\frac 16\frac{q^3}{Q^3}\right)\,,\\
\Gamma_f(q) &=& \frac{n_f}{3\pi}\frac{\as(q^2)}{q}\left(
1-\frac 32\frac{Q_0}{q} +\frac 32\frac{Q_0^2}{q^2}
-\frac{Q_0^3}{q^3}\right)\,.
\eeqn

The Sudakov factors for no resolvable emission are now
\beqn
\Delta_q(Q) &=& \exp\left(-\int_{Q_0}^{Q} dq\,\Gamma_q(Q,q)\right)\,,\\
\Delta_q(Q) &=& \exp\left(-\int_{Q_0}^{Q} dq\left[\Gamma_g(Q,q)
+\Gamma_f(q)\right]\right)\,.
\eeqn
Hence the rates for 1, 2 or 3 subjets in a quark jet are:
\beqn\label{eq:subjets_qrk}
R^q_1 &=& \Delta_q(Q)\,,\nn
R^q_2 &=& \Delta_q(Q)\int_{Q_0}^Q dq\,\Gamma_q(Q,q)\Delta_g(q)\,,\nn
R^q_3 &=& \Delta_q(Q)\int_{Q_0}^Q dq\int_{Q_0}^q
dq'\,\Gamma_q(Q,q)\Delta_g(q)\times\nn
&&\left\{\left[\Gamma_q(Q,q')+\Gamma_g(q,q')\right]\Delta_g(q')
+\Gamma_f(q')\Delta_f(q')\right\}\,,
\eeqn
where $\Delta_f=\Delta_q^2/\Delta_g$, and for a gluon jet we have
\beqn\label{eq:subjets_glu}
R^g_1 &=& \Delta_g(Q)\,,\nn
R^g_2 &=& \Delta_g(Q)\int_{Q_0}^Q
  dq\left[\Gamma_g(Q,q)\Delta_g(q)+\Gamma_f(q)\Delta_f(q)\right]\,,\nn
R^g_3 &=& \Delta_g(Q)\int_{Q_0}^Q dq\int_{Q_0}^q
dq'\biggl[\Gamma_g(Q,q)\Delta_g(q)\times\nn
&&\left\{\left[\Gamma_g(Q,q')+\Gamma_g(q,q')\right]\Delta_g(q')
+\Gamma_f(q')\Delta_f(q')\right\} +\Gamma_f(q)\Delta_f(q)\times\nn
&&\left\{\left[\Gamma_g(Q,q')-\Gamma_g(q,q')\right]\Delta_g(q')
+2\Gamma_q(q,q')\Delta_q(q')\right\}\biggr]\,.
\eeqn

\section{Subjet rates in jets: comparison with Monte Carlo}
\label{sec:6}
We now compare the above results with Monte Carlo predictions.
MC samples of quark and gluon jets were prepared for the subjet analysis using the same setup as in the associated jet study in Sec.~\ref{sec:2}, however, detector effects and minimum $p_T$ cuts for the charged and neutral hadrons were not included for this analysis.  In this sense, our study of the subjet rates should be taken as illustrative, and we do not include the subjet rates in an MVA analysis in this paper.  As we shall see in the following, one needs to go down to at least $L=4$ to have some discrimination power. This corresponds to going down to $0.1$ for  $\Delta R$ resolution, which is the typical size of calorimeter cells, although the $\Delta R$ separation of subjets would be larger when the subjet $p_T$ is smaller compared to the primary jet $p_T$.  Therefore, in a  proper analysis, combining track and calorimeter information is essential, and a detailed experimental study is necessary, which is beyond the
scope of this paper.

\begin{figure}[htb!]
\centering 
\includegraphics[width=0.45\textwidth]{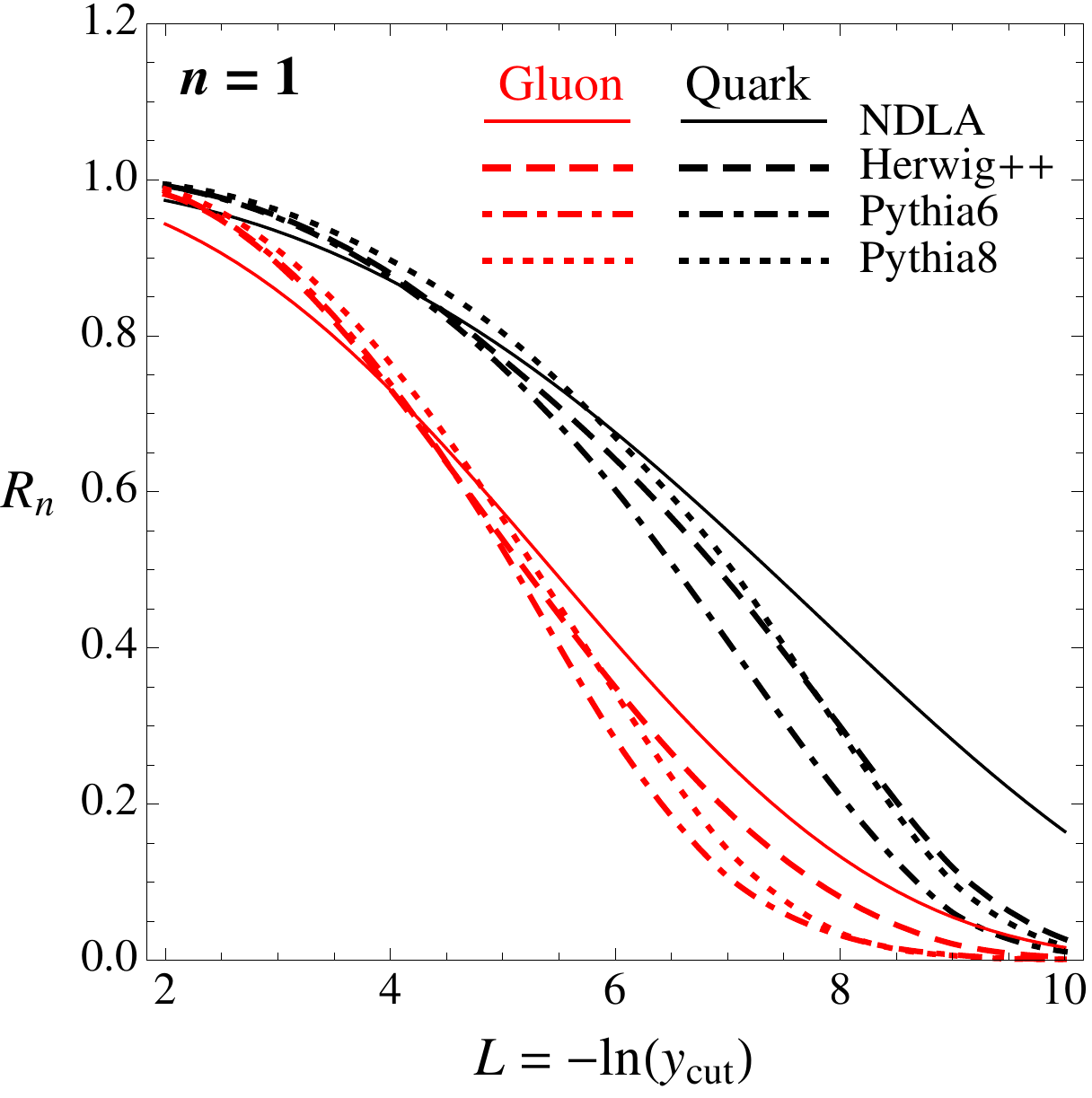}
\hfill
\includegraphics[width=0.45\textwidth]{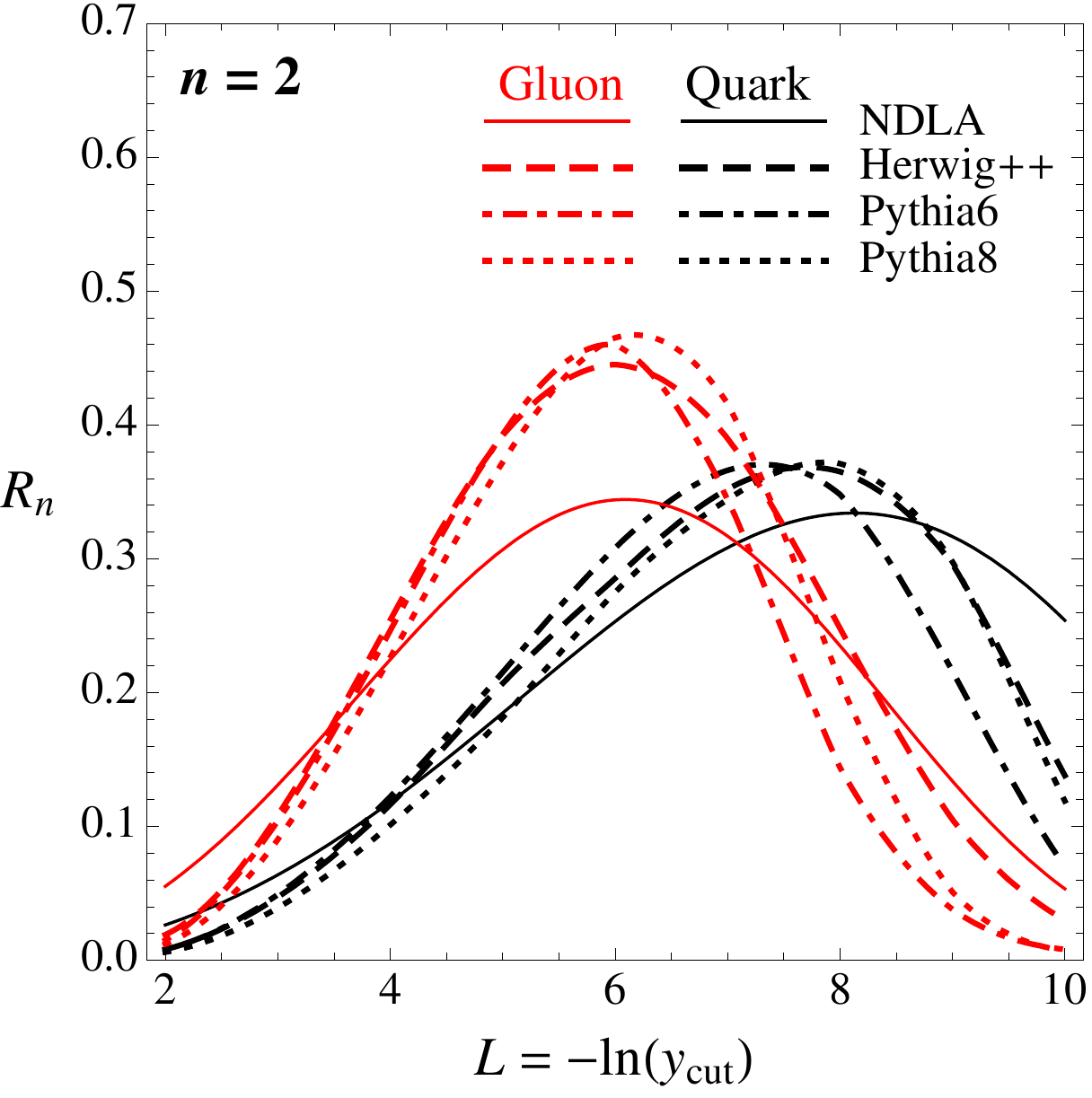}
\includegraphics[width=0.45\textwidth]{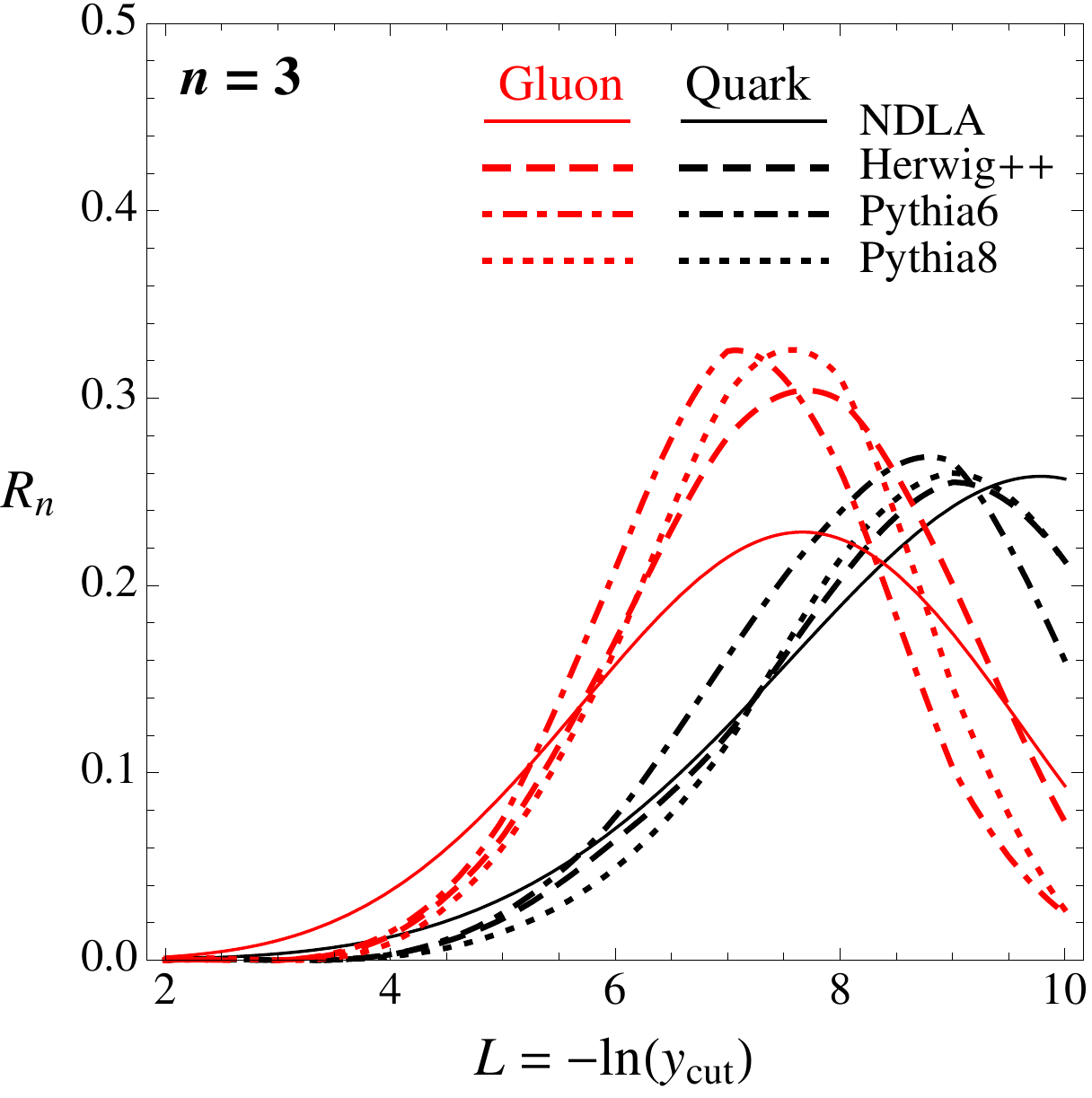}
\hfill
\includegraphics[width=0.45\textwidth]{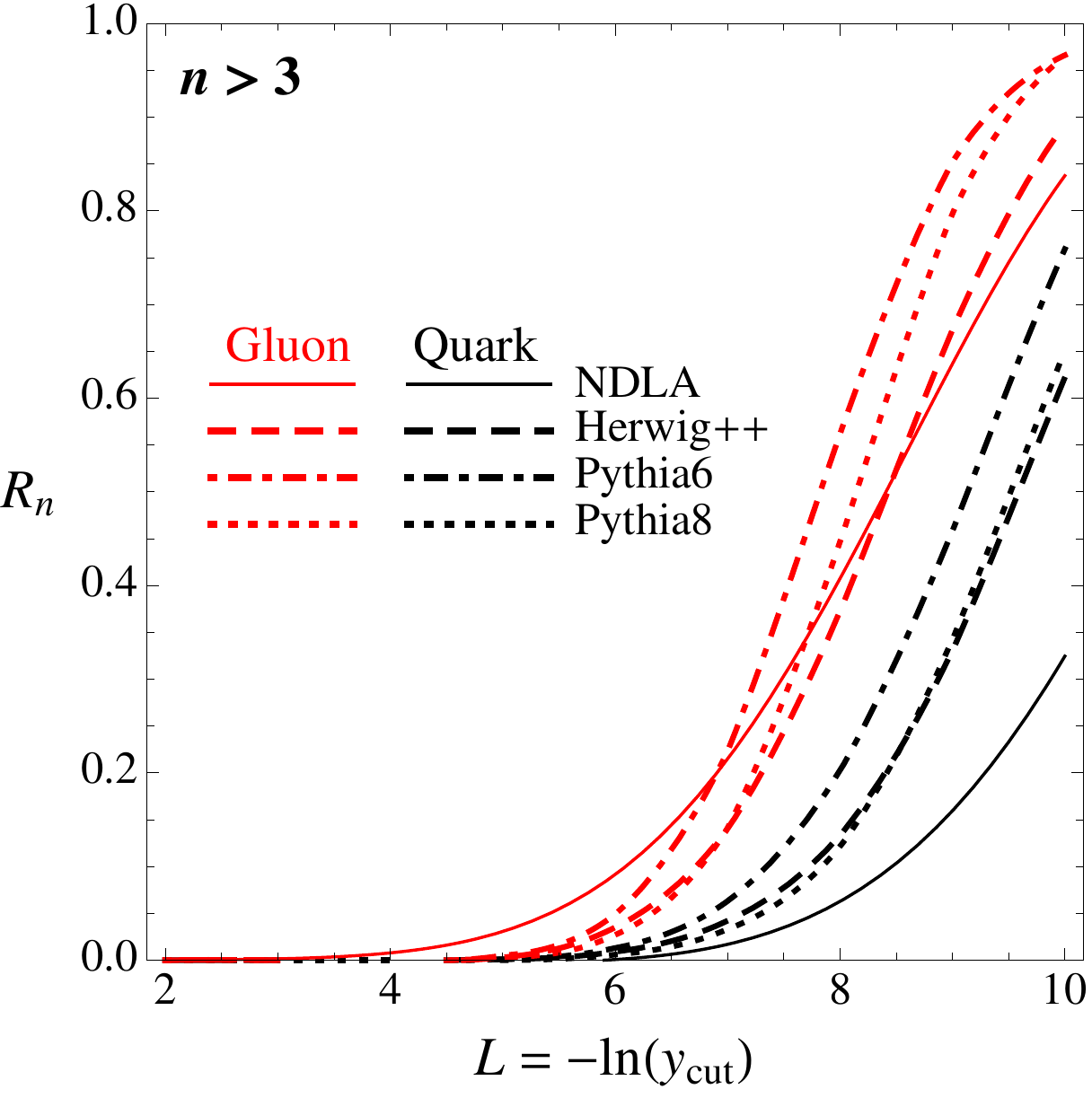}
\caption{\label{fig:5} Subjet rates $R_n$ with $n=1,2,3$ and $n >3$ as a function of $L=-{\rm ln}(y_{\rm cut})$, for quark jets (black) and gluon jets (red),
with $p_{T,J} \in [500,600]$ GeV, $R=0.4$. Curves are {\tt Herwig++} (dashed),
{\tt Pythia6} (dot-dashed), {\tt Pythia8} (dotted) and NDLA resummed (solid).}
\end{figure}

Figure~\ref{fig:5} shows comparisons between the resummed
results of Eqs.~(\ref{eq:subjets_qrk},\ref{eq:subjets_glu})
and the MC results for jets with $p_{T,J} \in [500,600]$ GeV and
$R=0.4$.  For quark jets the different MC's 
 agree quite well with each other and with the
resummed calculations, the MC predictions being somewhat below the resummed 1-subjet rate for $L>4$,
and vice-versa for 2 subjets.  Hadronization effects are small for $L<7$, after which the
1- and 2-subjet rates are suppressed and the higher subjet rates are
therefore enhanced.  At this value of $R\,p_{Tj}$, $L=7$ corresponds
to resolving subjets with $\min\{p_{ti},p_{tj}\}\Delta R_{ij}\sim 6$ GeV.

For gluon jets the agreement between the resummed
results and the Monte Carlos is still
quite close for 1 subjet.  For 2 and 3 subjets the peak rates are in 
roughly the same place but have higher values than the resummed ones,
with the effect that the rate for 4 or more subjets is substantially suppressed.
Once again the hadronization effects are small for $L<7$, after which the
1- and 2-subjet rates are suppressed and the higher subjet rates are
enhanced, actually bringing the latter into close agreement with the
analytical calculations.   

In conclusion, the fairly good agreement between the Monte Carlos and
the resummed 1-, 2- and 3-subjet rates for $R=0.4$ and $L$ not too
large ($L<5$, subjet resolution above about 15 GeV) suggests
that in this range those subjet rates can be used for quark-gluon
discrimination.  At larger jet radii, the agreement remains similar,
as we have checked using $R=0.8$. 

\section{Summary}
\label{sec:7}
To summarize our findings, we show that in studies of light quark and gluon jet separation at the LHC, it is important to include the information on associated jet rates around a primary hard jet. Associated jet rates are defined as the probability of finding at least one softer reconstructed jet around the primary hard jet under consideration. This probability is found to be substantially higher for a gluon-initiated jet compared to a quark-initiated one. Since commonly a small jet radius parameter is adopted in LHC studies of hadronic jets, the associated jet rates carry the information on the radiation outside the chosen jet radius. 

We compute the associated jet rates up to NDLA accuracy in perturbative QCD, as a function of the primary jet and minimum associated jet $p_T$'s, as well as the jet radius and association radius parameters. The NDLA results are thereafter compared with predictions from different parton shower MC's. Since the NDLA predictions include only the time-like showering of the final state partons, we demonstrate the effects of ISR and MPI in the MC predictions as well, and it is observed that the NDLA predictions are closer to the MC's when ISR and MPI are switched off. Overall, the associated jet rates are not very sensitive to these effects as long as the association radius is not too large. 

The probability of having at least one associated jet for a primary gluon jet is roughly a factor of two larger than for a quark jet, with a small variation in this number as a function of the jet $p_T$. This fact makes the presence or absence of associated jets a good variable for quark-gluon discrimination studies. We demonstrate the impact of including the associated jet rate information by including this variable in an MVA analysis, along with the well-studied variables of number of charged tracks, energy-energy-correlation angularities and jet mass. Comparing different two and three variable MVA's with and without the associated jet information, we find that including the associated jets leads to an improvement of around $10\%$ in rejecting gluons, for a fixed quark selection efficiency of $0.4$. We also show that using a three variable MVA with associated jet categories leads to the best performance, with an improvement of $20\%$ in rejecting gluons, for the same quark efficiency as above. 

Since for the number of charged tracks variable the MC predictions tend to differ, and are dependent on the parton shower and underlying event parameter tunes, we explore the number of $k_t$ subjets of an anti-$k_t$ jet as a quark-gluon separation variable. We compute the number of subjets to NDLA accuracy, and compare the resummed predictions with different MC's. The different MC predictions are found to be rather uniform, with the resummed predictions being broadly in agreement with them. However, for gluon jets the peak rates for 2 and 3 subjets are found to be lower in the resummed computation, which might arise due to higher-order effects that are in general bigger for gluons.

\section*{Acknowledgments}
This work is supported by JSPS KAKENHI No. 26287039 and Grant- in-Aid for Scientific research from the Ministry of Education, Science, Sports, and Culture (MEXT), Japan, No. 23104006, and also by the World Premier International Research Center Initiative (WPI Initiative), MEXT, Japan. B.~R.~Webber thanks the Kavli IPMU for hospitality during part of this work.

\appendix
\section{Distributions of discrimination variables}
\label{appx}
\begin{figure}[htb!]
\centering 
\includegraphics[keepaspectratio=true, scale = 0.58]{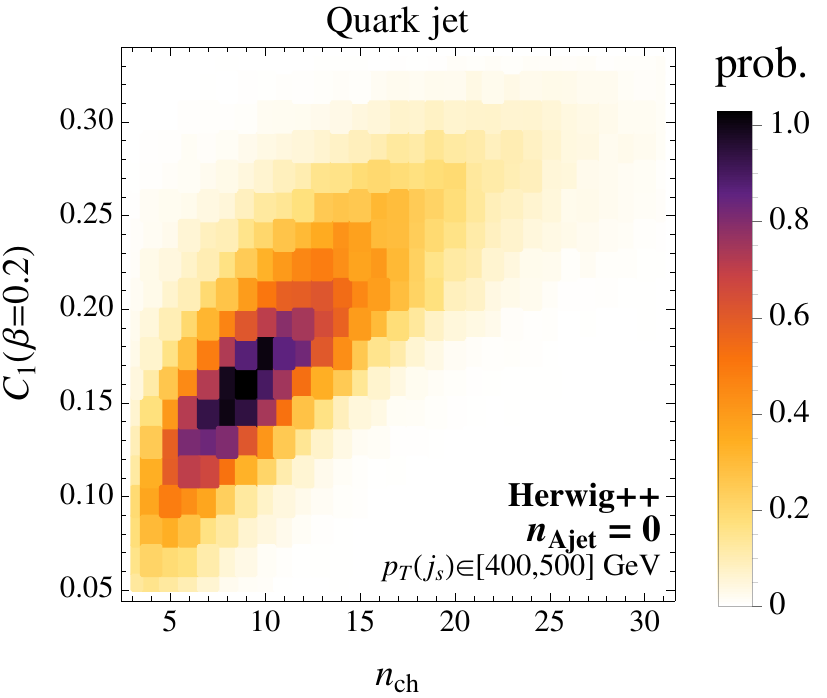}
\includegraphics[keepaspectratio=true, scale = 0.58]{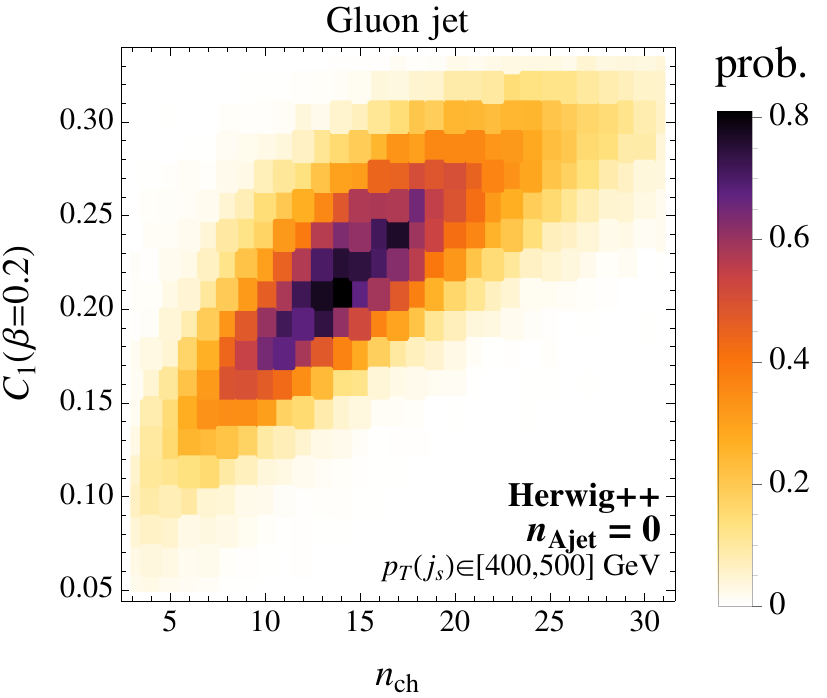}
\includegraphics[keepaspectratio=true, scale = 0.58]{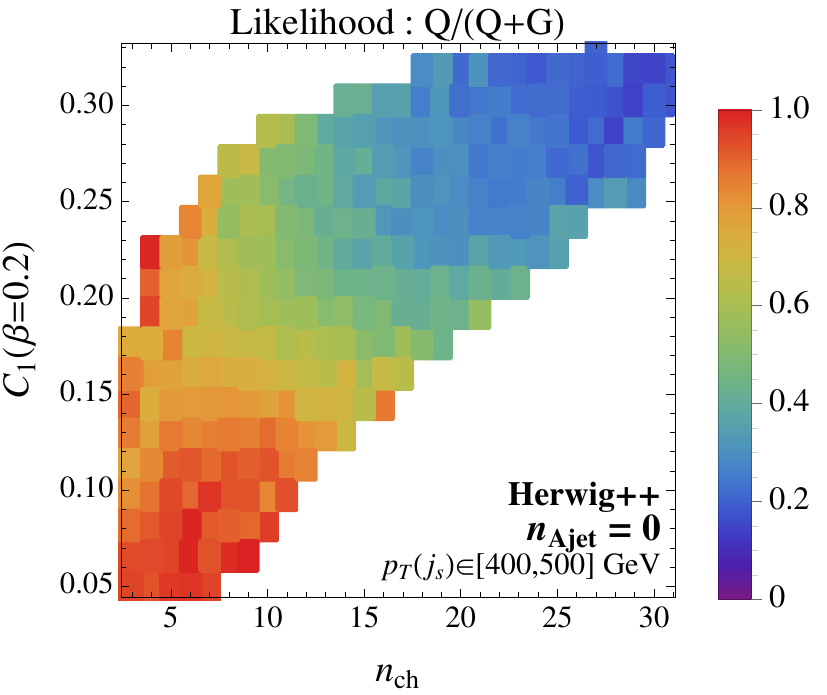}
\vfill
\includegraphics[keepaspectratio=true, scale = 0.58]{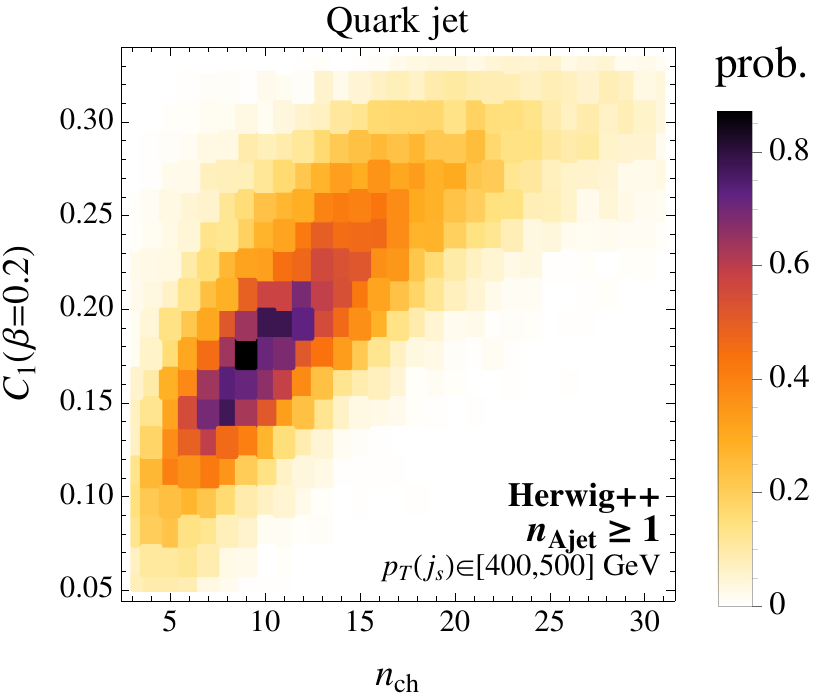}
\includegraphics[keepaspectratio=true, scale = 0.58]{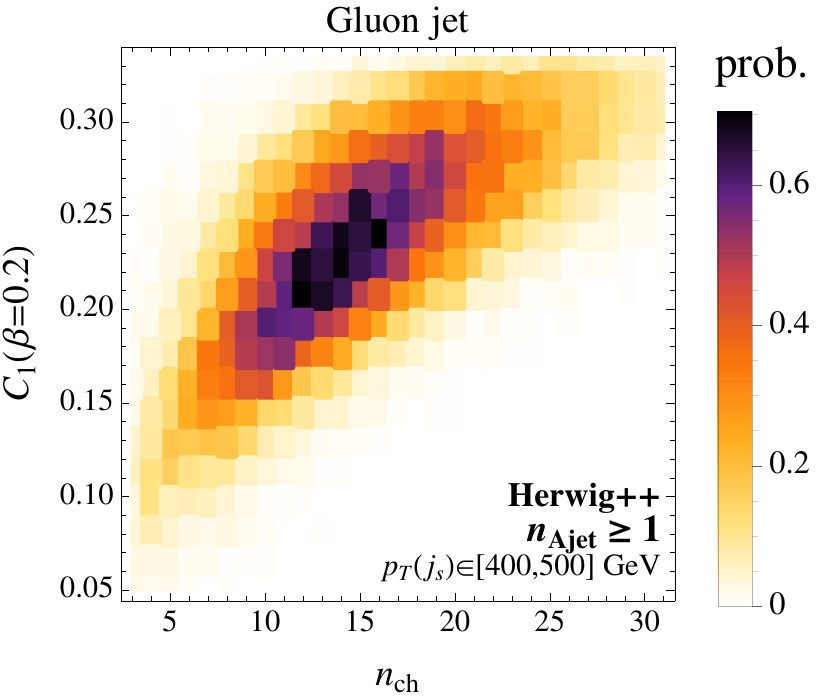}
\includegraphics[keepaspectratio=true, scale = 0.58]{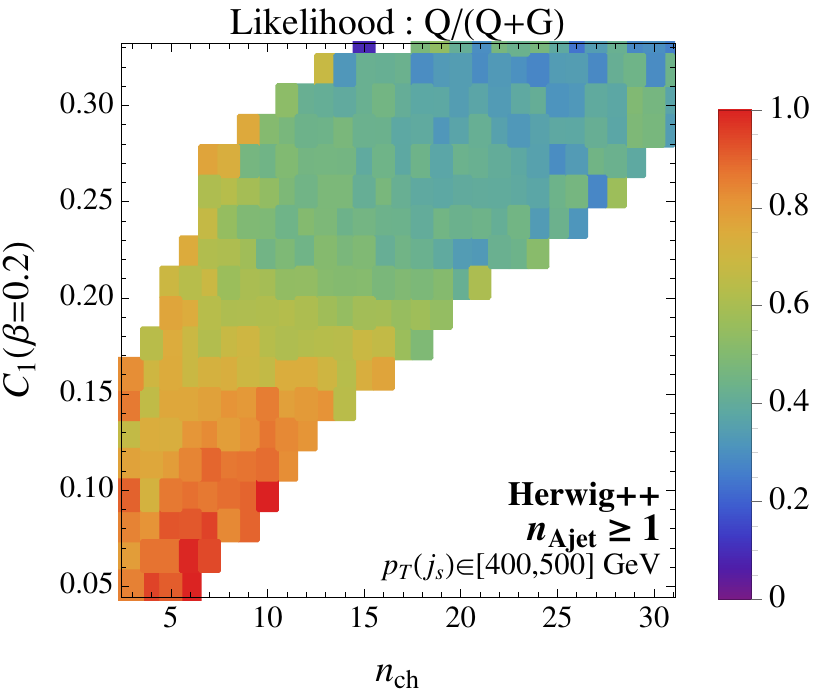}
\includegraphics[keepaspectratio=true, scale = 0.58]{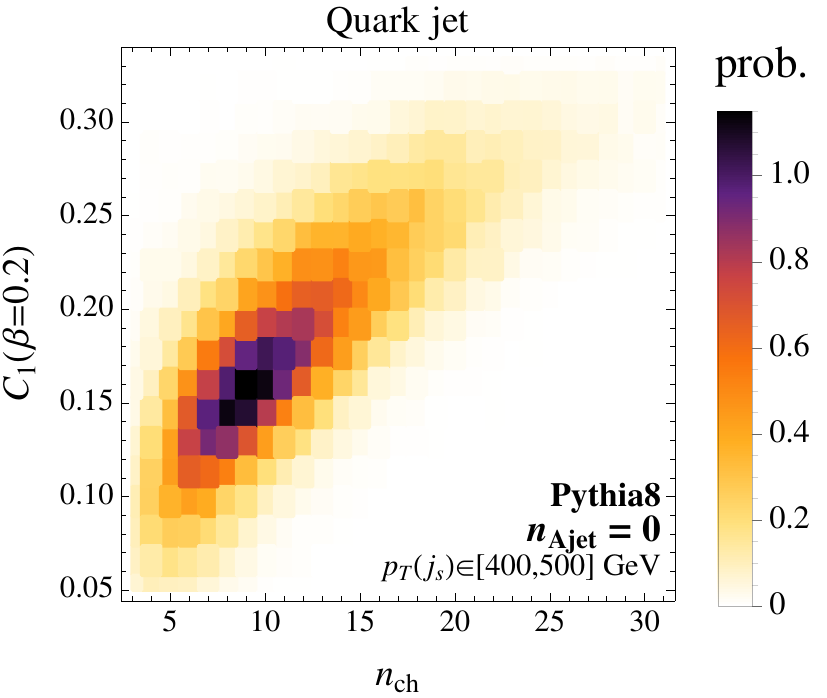}
\includegraphics[keepaspectratio=true, scale = 0.58]{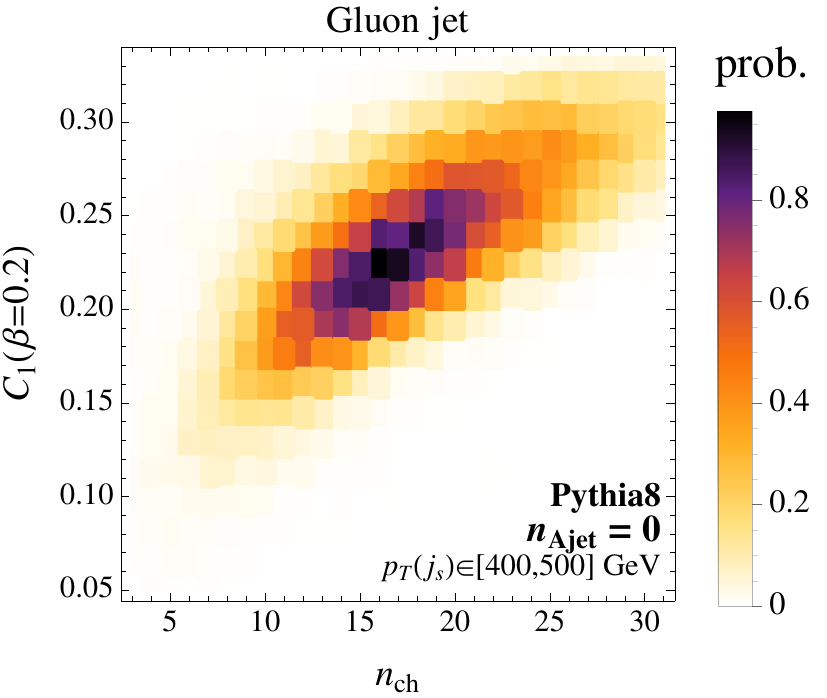}
\includegraphics[keepaspectratio=true, scale = 0.58]{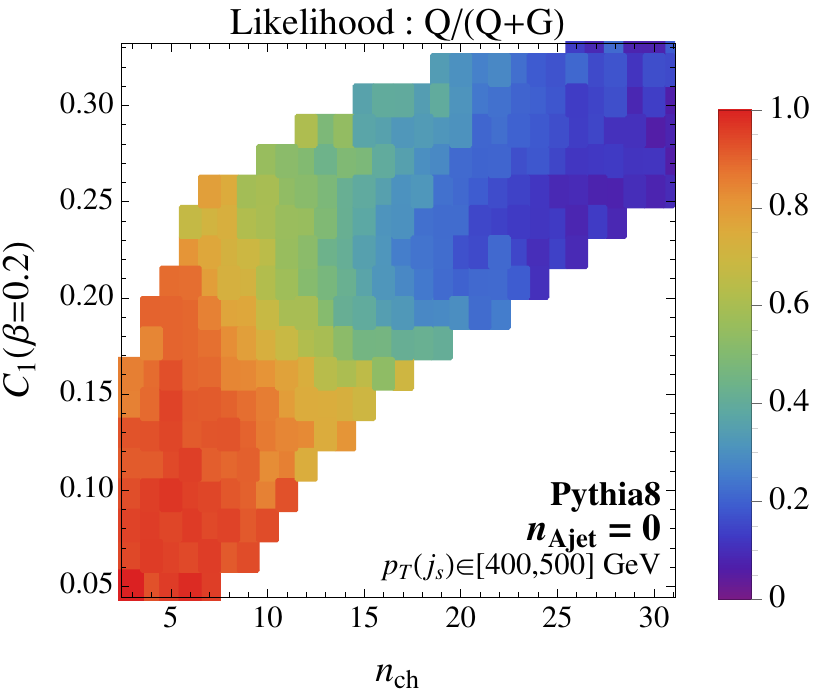}
\vfill
\includegraphics[keepaspectratio=true, scale = 0.58]{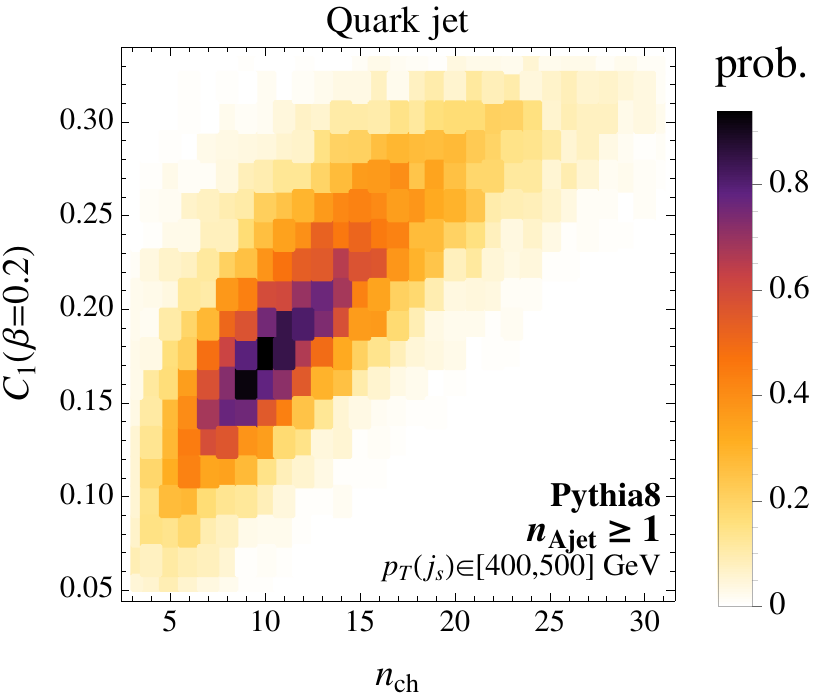}
\includegraphics[keepaspectratio=true, scale = 0.58]{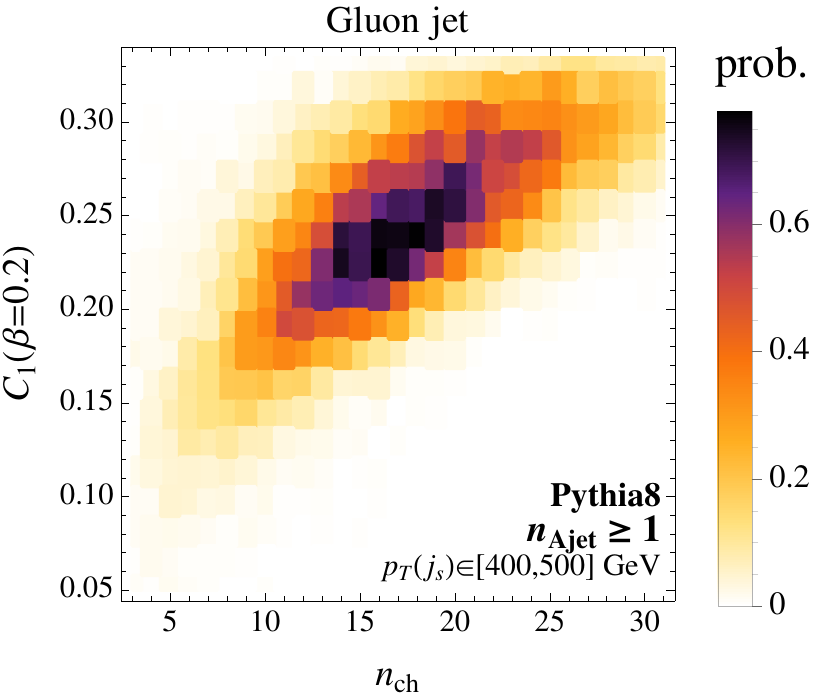}
\includegraphics[keepaspectratio=true, scale = 0.58]{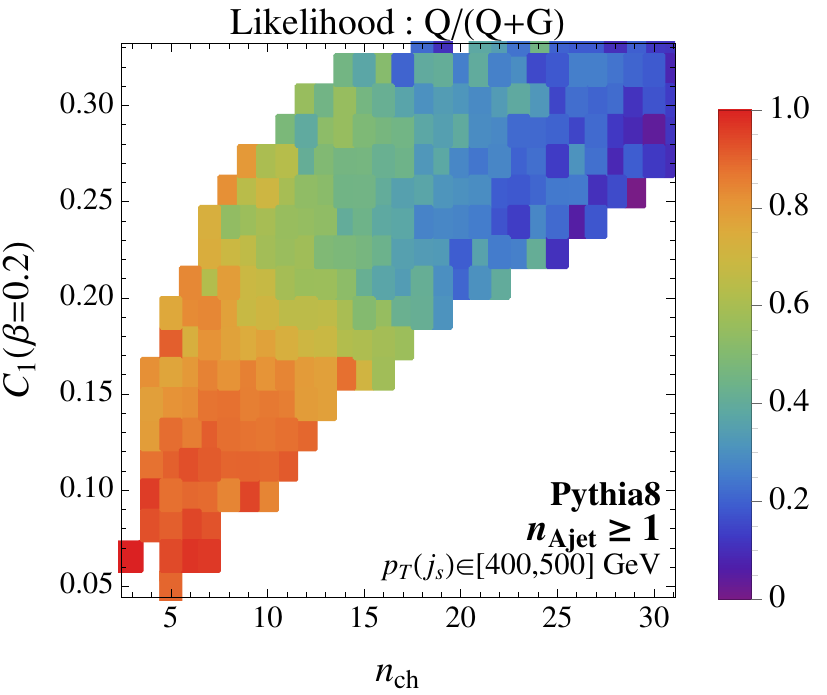}
\caption{\label{fig:Comb1} Joint distributions of $n_{\rm ch}$ and $C_1^{(\beta=0.2)}$ in  {\tt Herwig++} and {\tt Pythia8}, for quark and gluon jets with $p_T(j_s)\in [400,500]$ GeV having $n_{\rm Ajet}=0$ and $\geq 1$ associated jets.}
\end{figure}

\begin{figure}[htb!]
\centering 
\includegraphics[keepaspectratio=true, scale = 0.58]{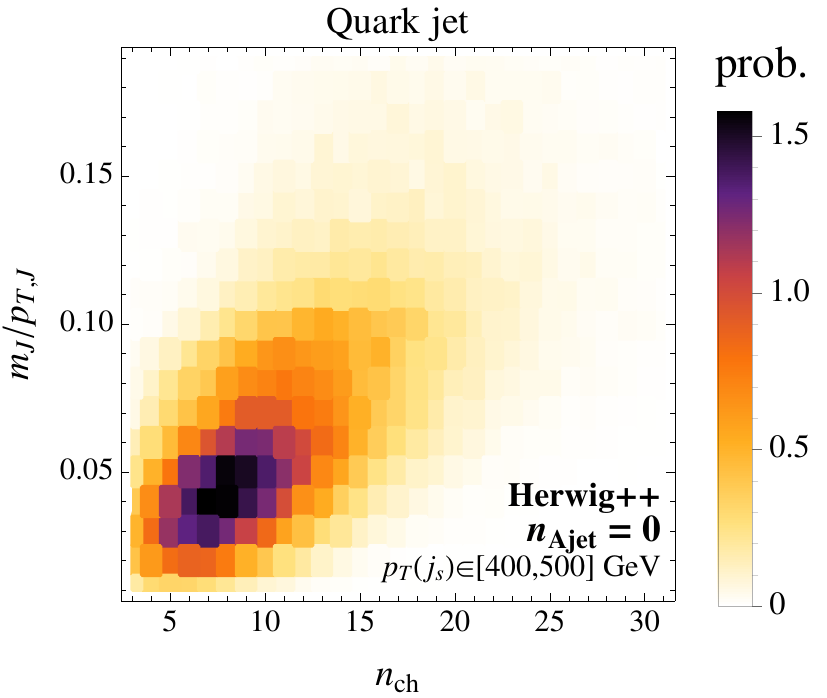}
\includegraphics[keepaspectratio=true, scale = 0.58]{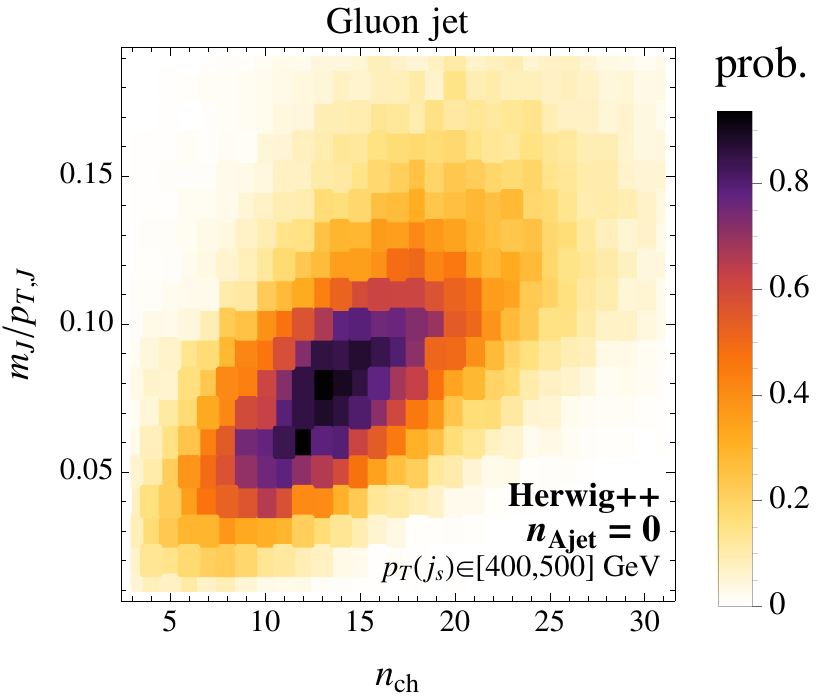}
\includegraphics[keepaspectratio=true, scale = 0.58]{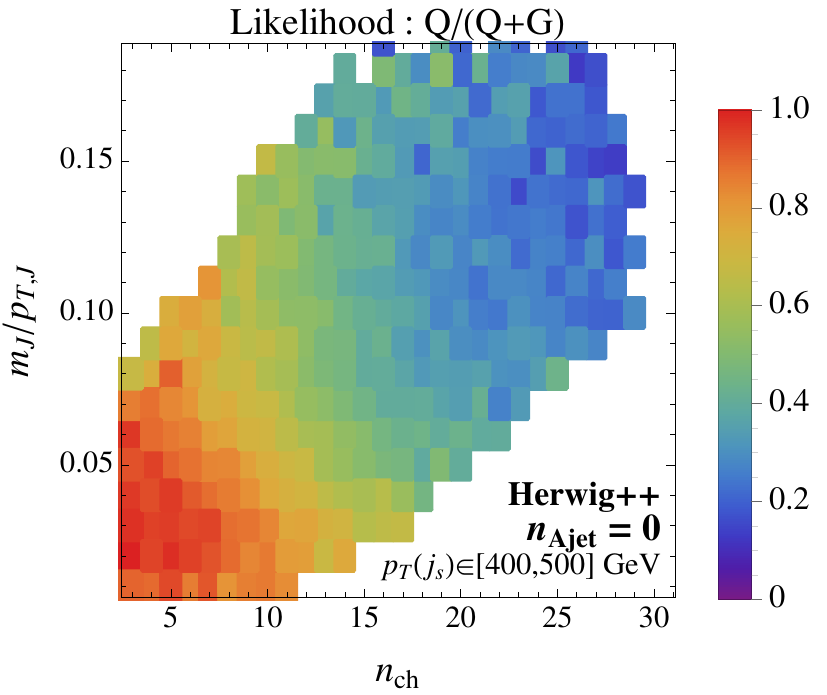}
\vfill
\includegraphics[keepaspectratio=true, scale = 0.58]{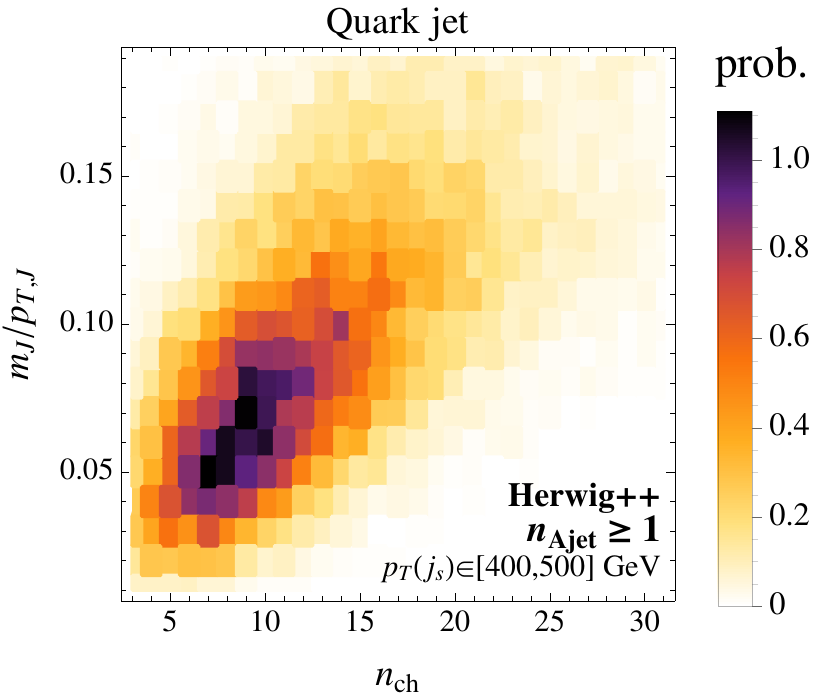}
\includegraphics[keepaspectratio=true, scale = 0.58]{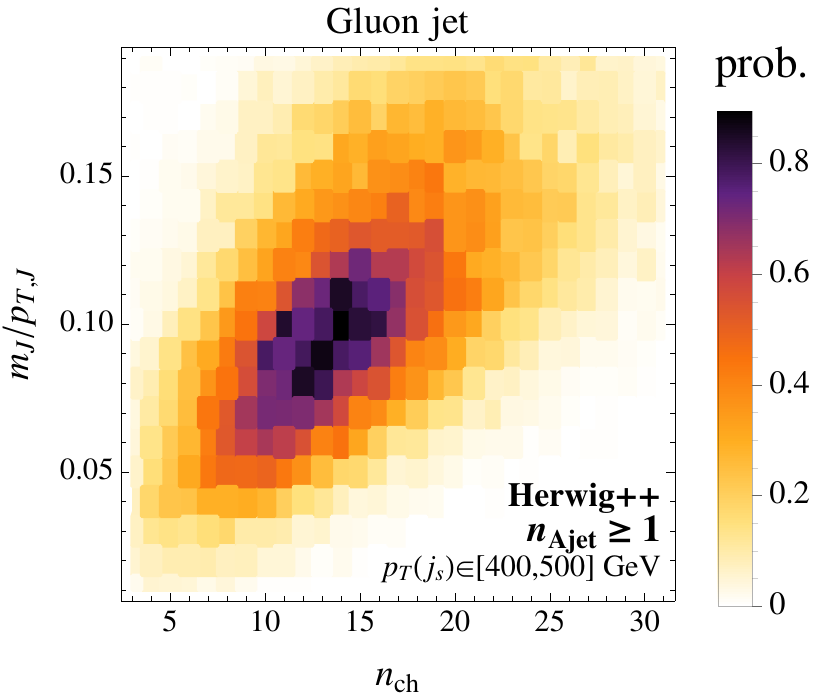}
\includegraphics[keepaspectratio=true, scale = 0.58]{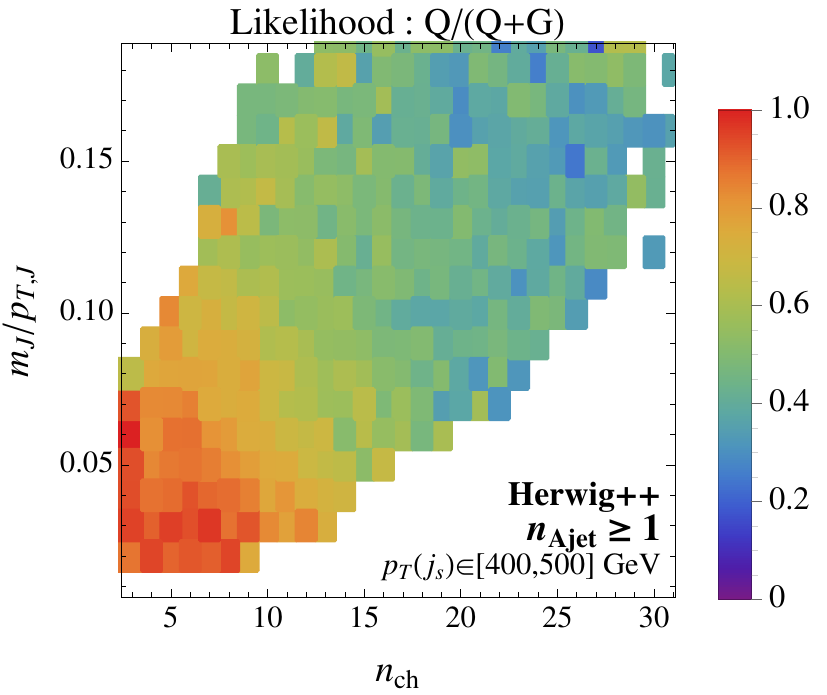}
\includegraphics[keepaspectratio=true, scale = 0.58]{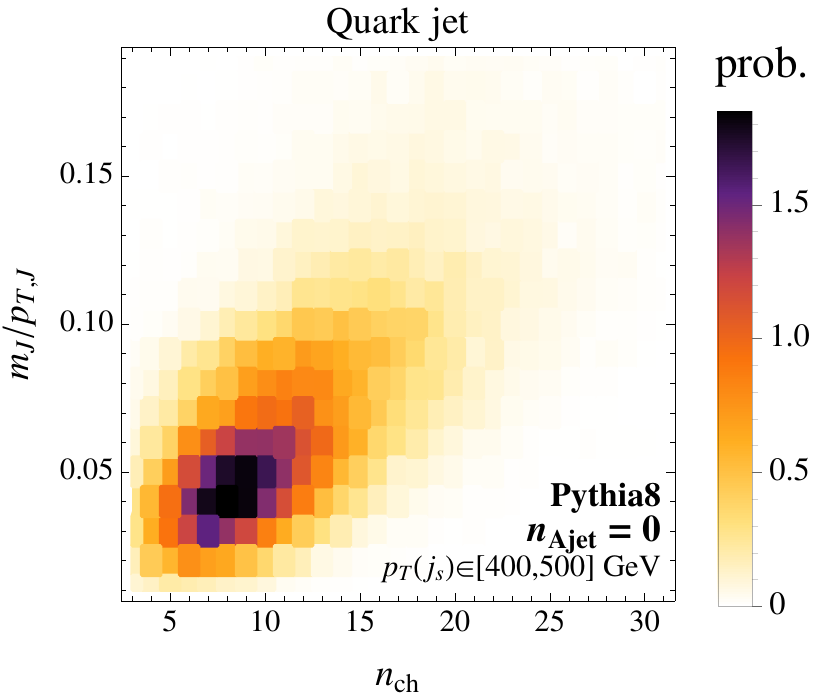}
\includegraphics[keepaspectratio=true, scale = 0.58]{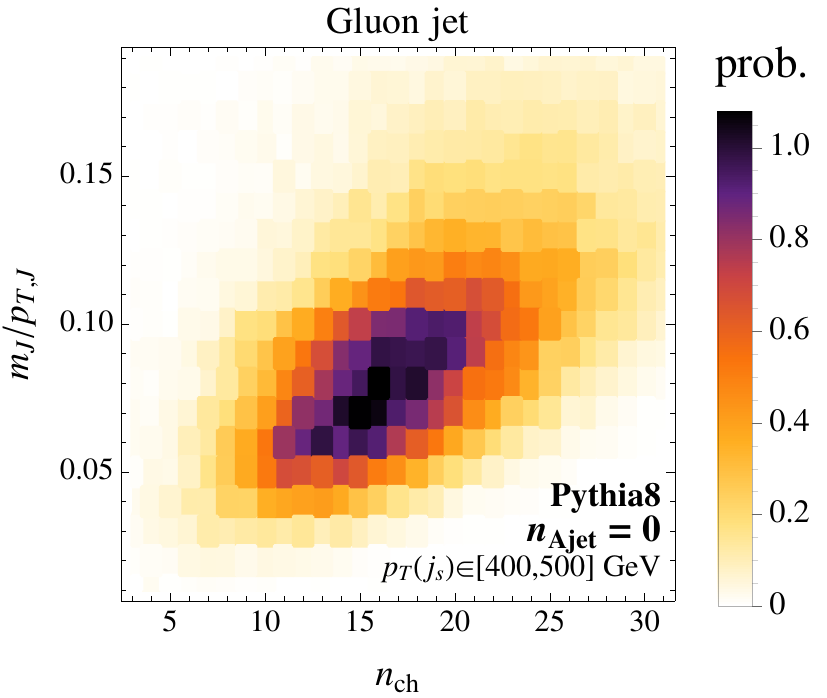}
\includegraphics[keepaspectratio=true, scale = 0.58]{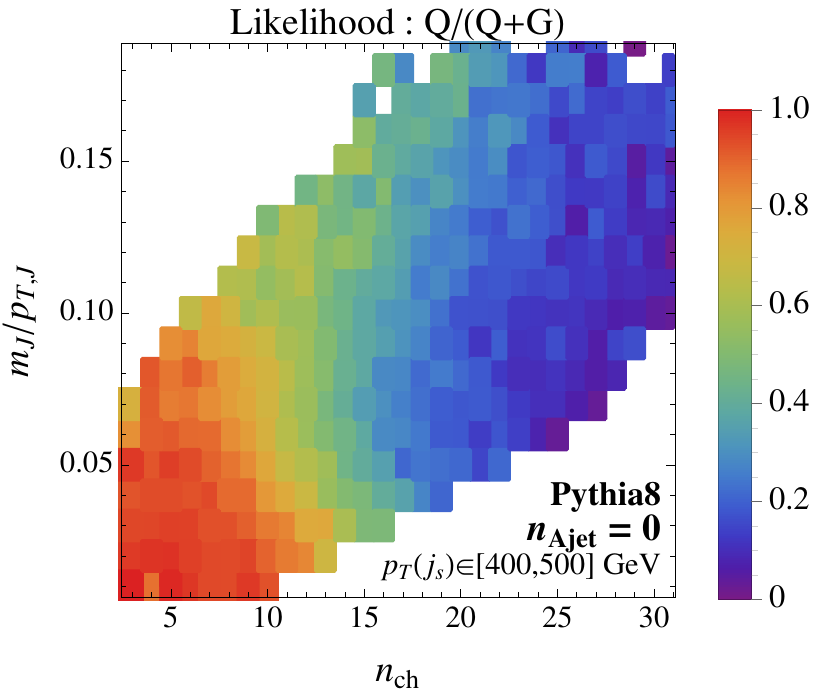}
\vfill
\includegraphics[keepaspectratio=true, scale = 0.58]{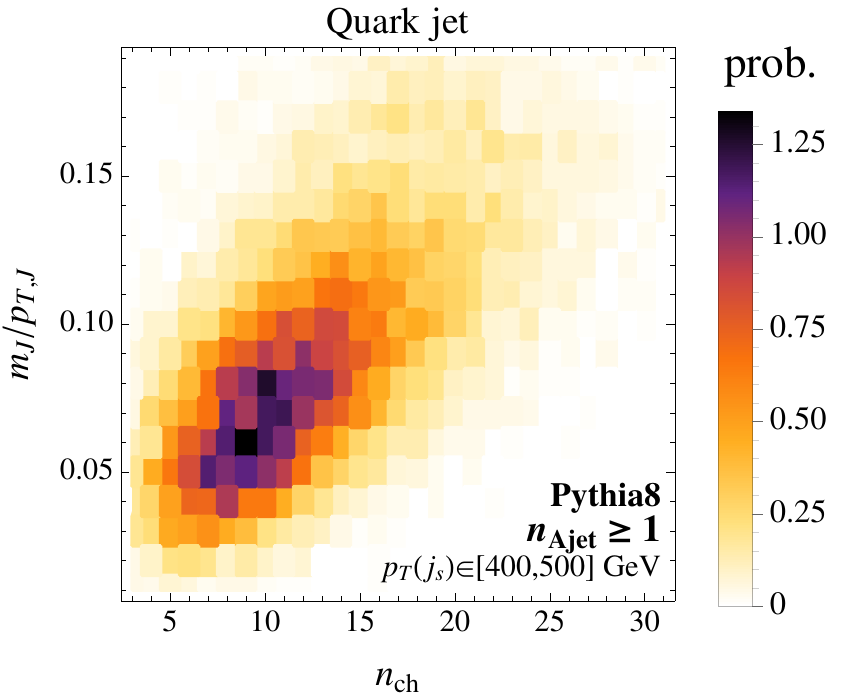}
\includegraphics[keepaspectratio=true, scale = 0.58]{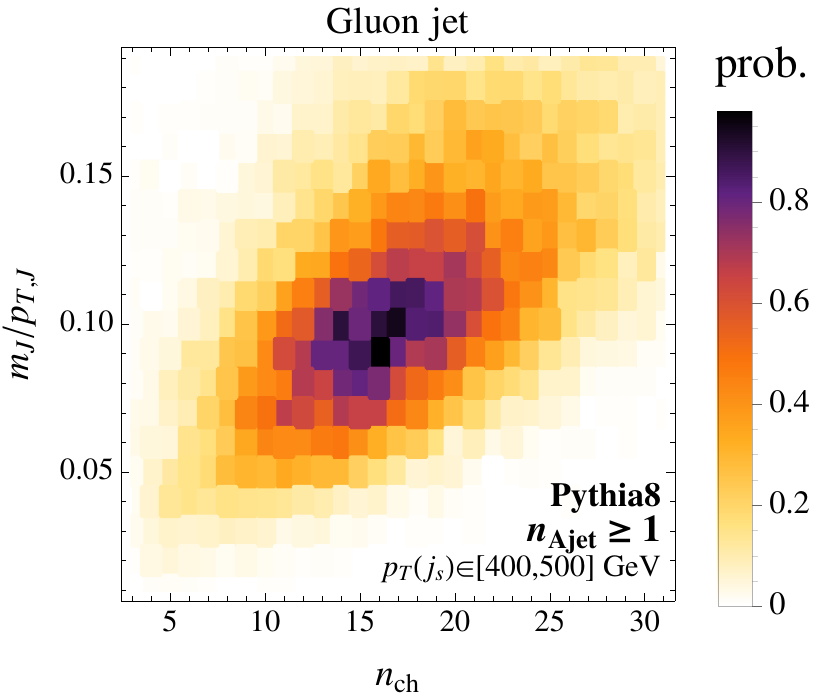}
\includegraphics[keepaspectratio=true, scale = 0.58]{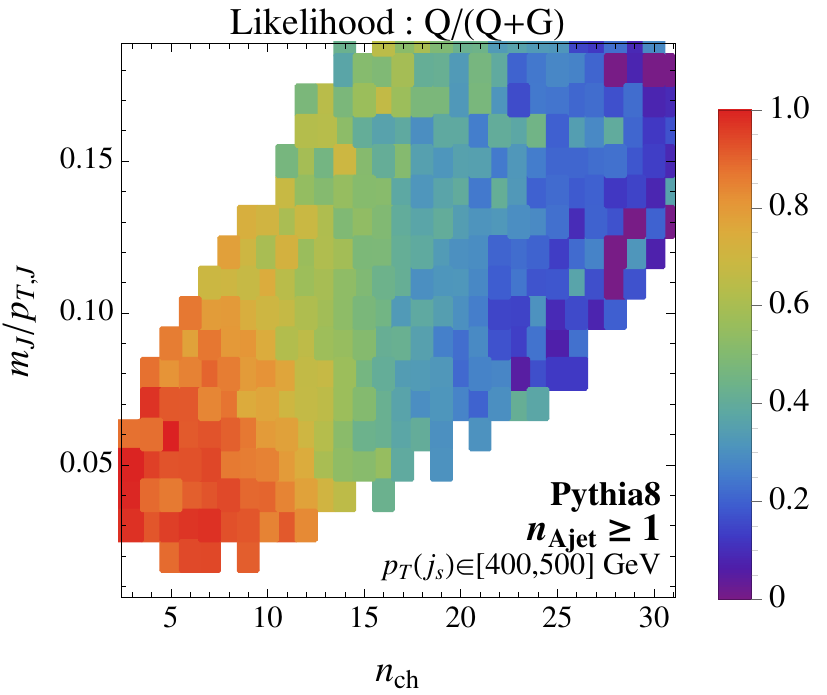}
\caption{\label{fig:Comb2} Joint distributions of $n_{\rm ch}$ and $m_J/p_{T,J}$
 in  {\tt Herwig++} and {\tt Pythia8}, for quark and gluon jets with $p_T(j_s)\in [400,500]$ GeV having $n_{\rm Ajet}=0$ and $\geq 1$ associated jets.}
\end{figure}

\begin{figure}[htb!]
\centering 
\includegraphics[keepaspectratio=true, scale = 0.58]{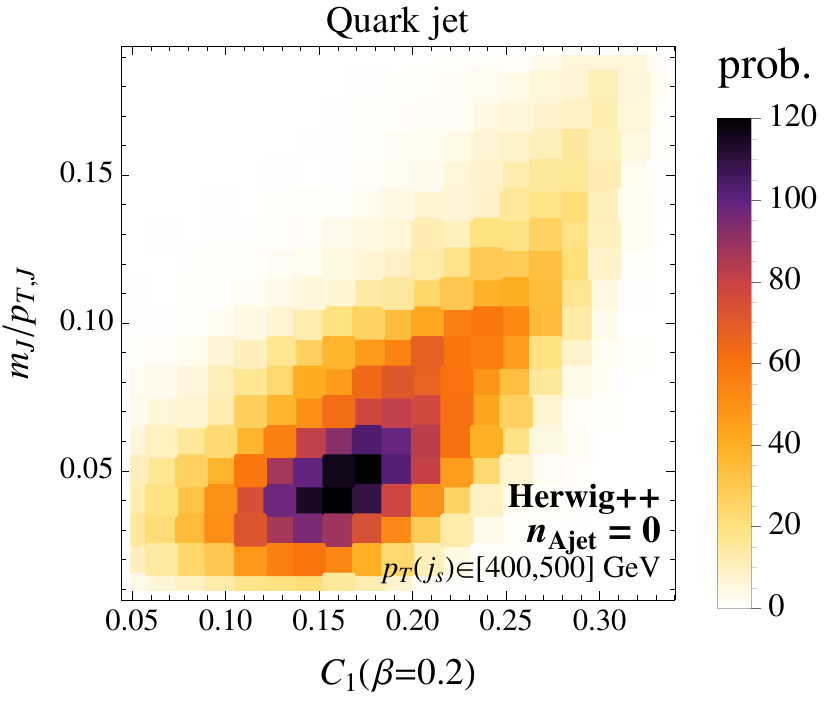}
\includegraphics[keepaspectratio=true, scale = 0.58]{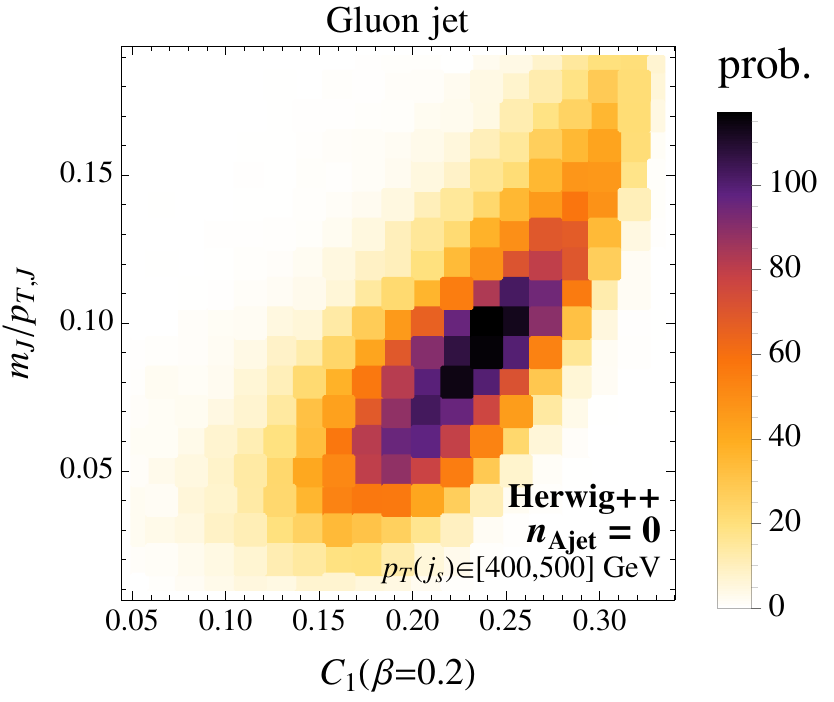}
\includegraphics[keepaspectratio=true, scale = 0.58]{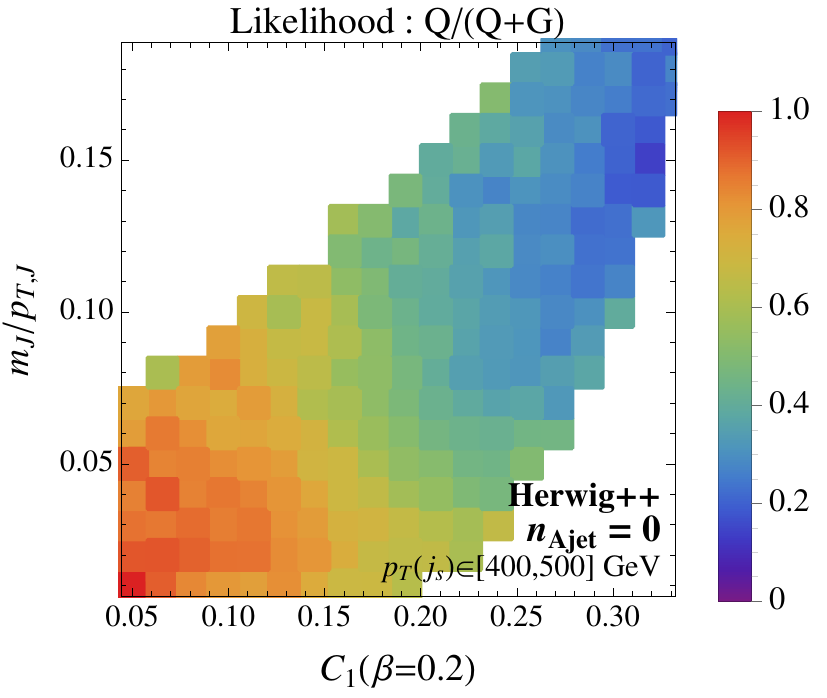}
\vfill
\includegraphics[keepaspectratio=true, scale = 0.58]{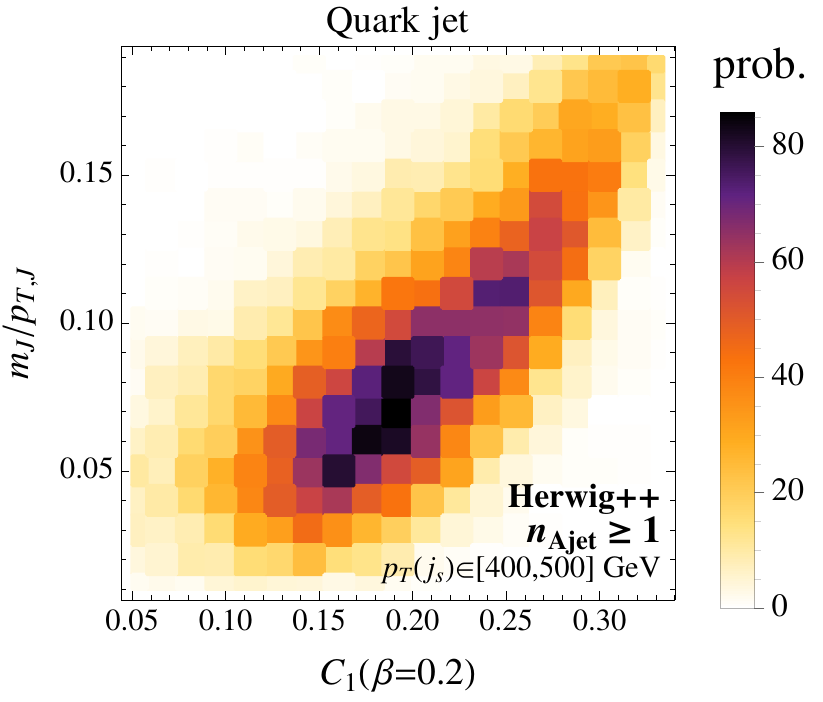}
\includegraphics[keepaspectratio=true, scale = 0.58]{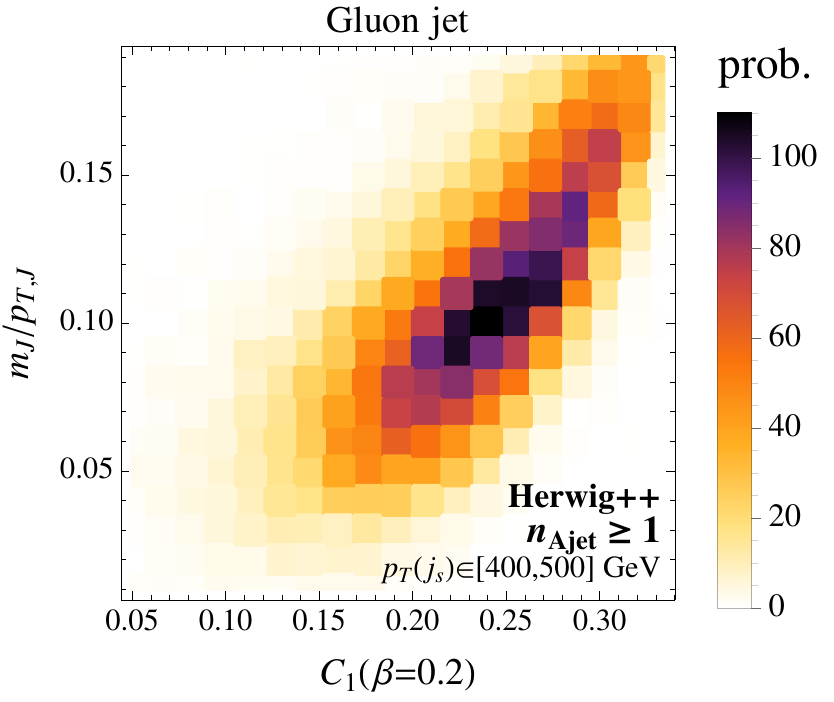}
\includegraphics[keepaspectratio=true, scale = 0.58]{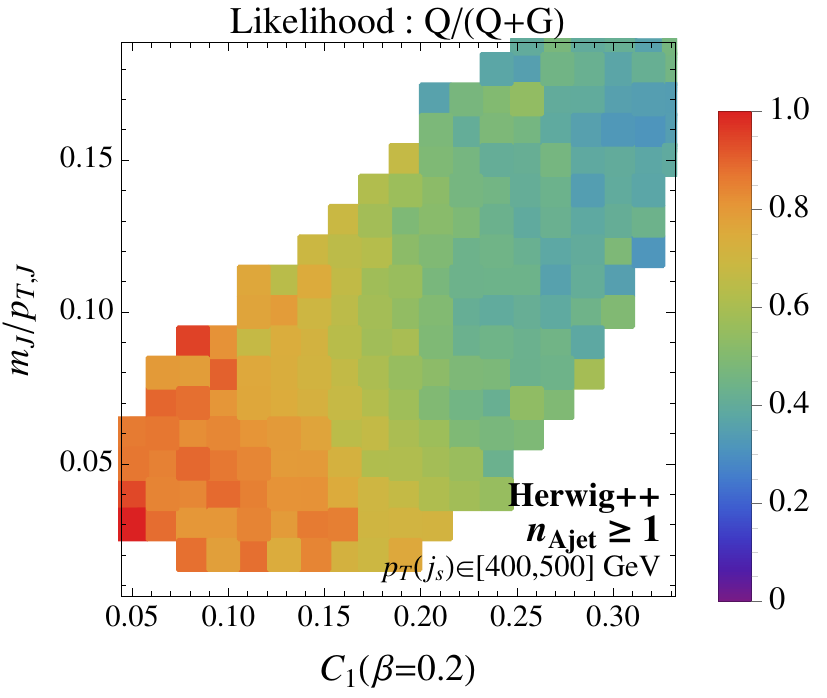}
\includegraphics[keepaspectratio=true, scale = 0.58]{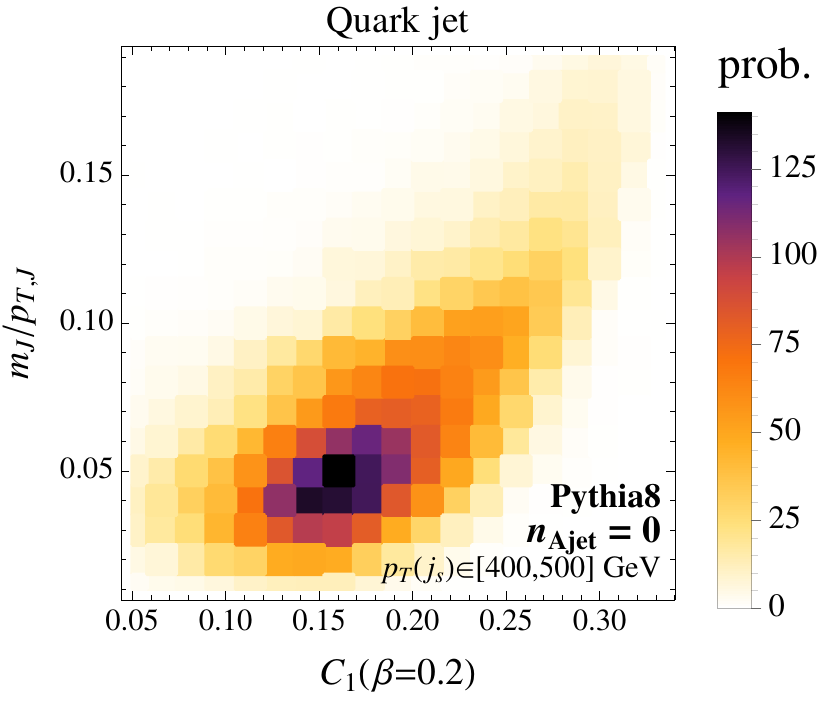}
\includegraphics[keepaspectratio=true, scale = 0.58]{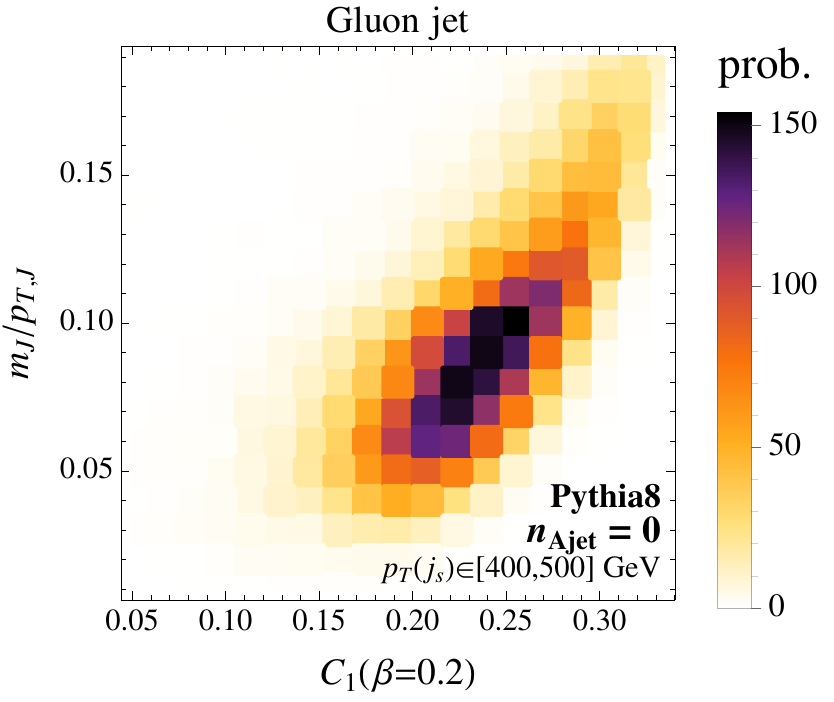}
\includegraphics[keepaspectratio=true, scale = 0.58]{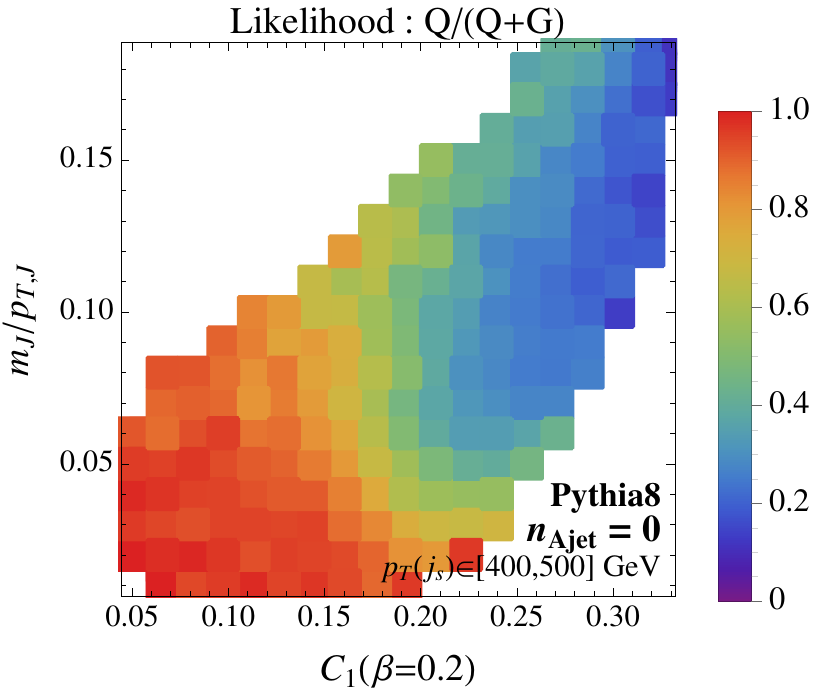}
\vfill
\includegraphics[keepaspectratio=true, scale = 0.58]{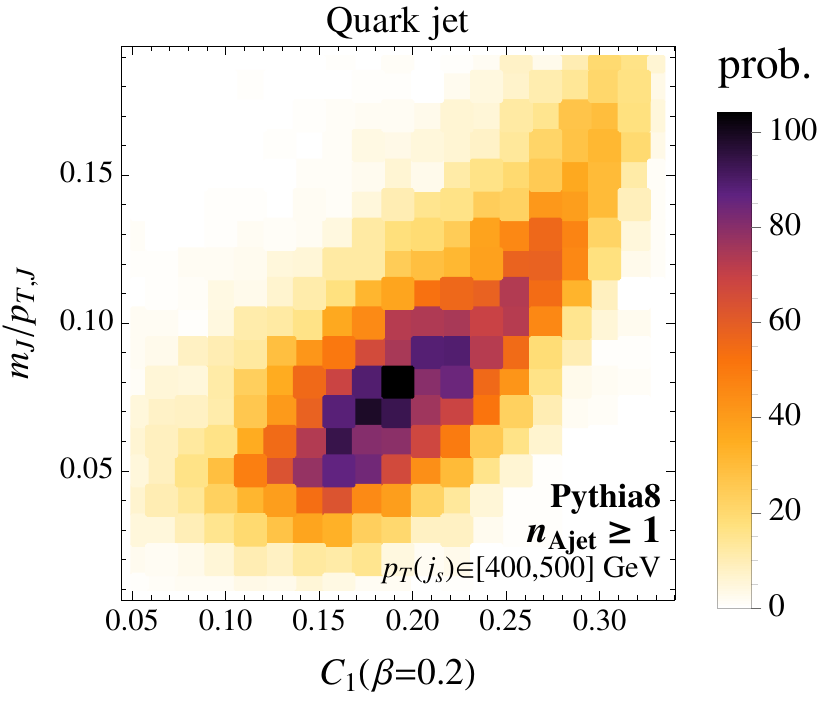}
\includegraphics[keepaspectratio=true, scale = 0.58]{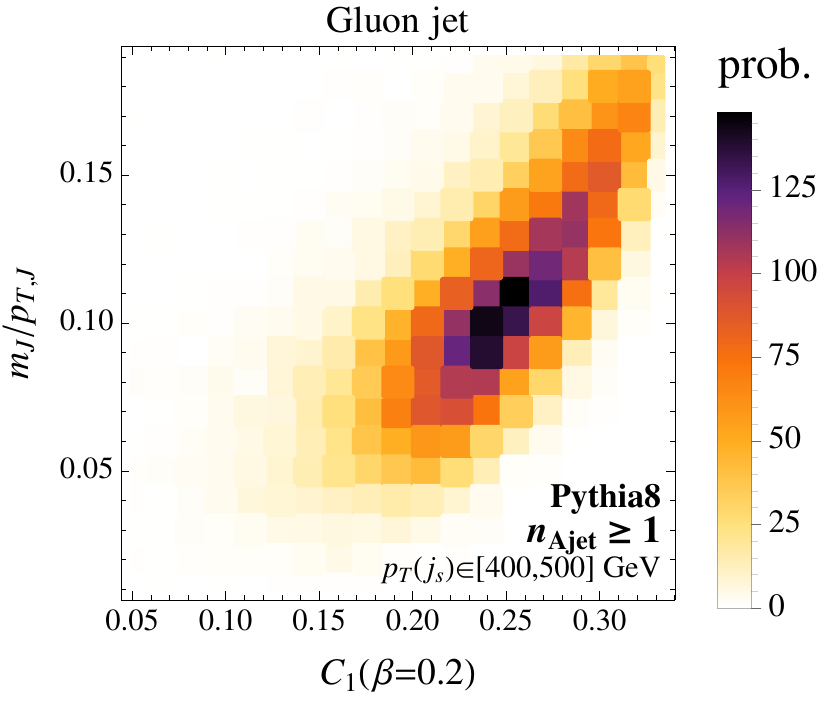}
\includegraphics[keepaspectratio=true, scale = 0.58]{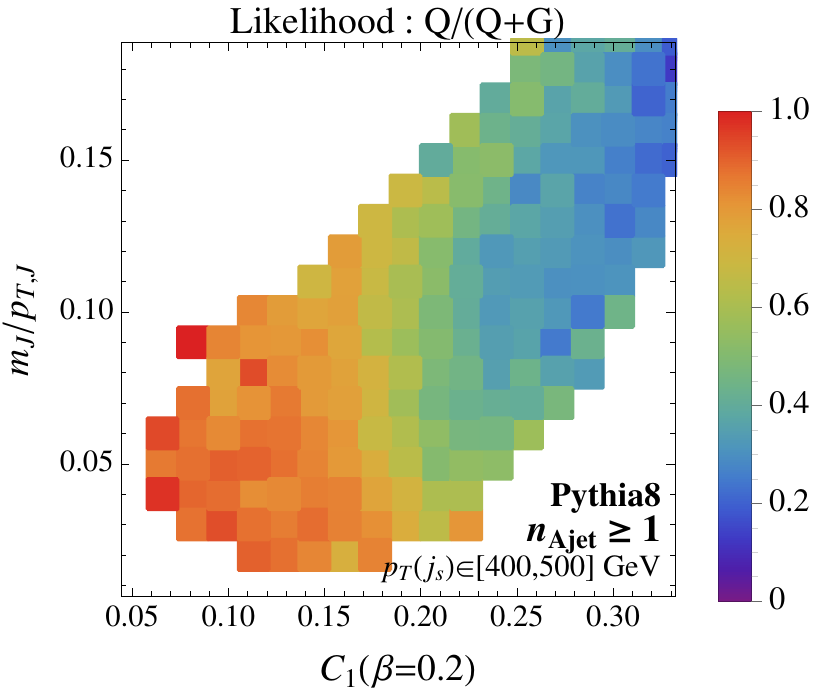}
\caption{\label{fig:Comb3} Joint distributions of $C_1^{(\beta=0.2)}$ and $m_J/p_{T,J}$ in  {\tt Herwig++} and {\tt Pythia8}, for quark and gluon jets with $p_T(j_s)\in [400,500]$ GeV having $n_{\rm Ajet}=0$ and $\geq 1$ associated jets.}
\end{figure}
In Figs.~\ref{fig:Comb1}-\ref{fig:Comb3} we show 2-dimensional plots of the joint distributions of the three discrimination variables used in the MVA presented in Section~\ref{sec:4}, for the two Monte Carlo event generators   {\tt Herwig++}
and {\tt Pythia8}.   The following features may be observed:
\begin{itemize}
\item
There are differences between the distributions predicted by the two Monte Carlos,
those of {\tt Pythia8} being somewhat narrower for quark jets and substantially
narrower for gluon jets.
\item
The distributions of the infrared-unsafe variable $n_{\rm ch}$ show the greatest
differences, with those of {\tt Pythia8}  being larger at high $n_{\rm ch}$.  This
could be due to differences in tuning of the non-perturbative parameters
of the generators.
\item
The above features are reflected in the likelihood plots, showing the probability
ratio $P_q/(P_q+P_g)$, and account for the higher discrimination efficiency
predicted by  {\tt Pythia8}  (Fig.~\ref{fig:3a} vs Fig.~\ref{fig:3}).
\item
The quark-gluon discrimination in the events with associated jets is weaker
than that for $n_{\rm Ajet}=0$.  This is expected because the events are
selected according to $p_T(j_s)$, the sum of leading and associated jet
$p_T$'s.  Therefore those with associated jets have leading jets with
lower $p_T$'s, which have lower discriminating power.
\item
Nevertheless the inclusion of the associated jet category
improves the MVA performance, because the probability of an associated
jet is lower for quark jets.
\end{itemize}



\begin{thebibliography}{99}

 \bibitem{Schwartz1}
  J.~Gallicchio and M.~D.~Schwartz,
  ``Quark and Gluon Tagging at the LHC,''
  Phys.\ Rev.\ Lett.\  {\bf 107} (2011) 172001.
 
 \bibitem{Schwartz2}
  J.~Gallicchio, J.~Huth, M.~Kagan, M.~D.~Schwartz, K.~Black and B.~Tweedie,
  ``Multivariate discrimination and the Higgs + W/Z search,''
  JHEP {\bf 1104} (2011) 069.
  
\bibitem{Schwartz3}
  J.~Gallicchio and M.~D.~Schwartz,
  ``Quark and Gluon Jet Substructure,''
  JHEP {\bf 1304} (2013) 090.

\bibitem{Larkoski1}
  A.~J.~Larkoski, G.~P.~Salam and J.~Thaler,
  ``Energy Correlation Functions for Jet Substructure,''
  JHEP {\bf 1306} (2013) 108.
  
\bibitem{Larkoski2}
  A.~J.~Larkoski, J.~Thaler and W.~J.~Waalewijn,
  ``Gaining (Mutual) Information about Quark/Gluon Discrimination,''
  JHEP {\bf 1411} (2014) 129.
  
\bibitem{ATLAS}
  G.~Aad {\it et al.}  [ATLAS Collaboration],
  ``Light-quark and gluon jet discrimination in $pp$ collisions at $\sqrt{s}=7\mathrm {\ TeV}$ with the ATLAS detector,''
  Eur.\ Phys.\ J.\ C {\bf 74} (2014) 8,  3023.
  
\bibitem{CMS}
CMS Collaboration, 
``Performance of quark/gluon discrimination using $pp$ collision data at $\sqrt{s}=8\mathrm {\ TeV}$,"
CMS-PAS-JME-13-002.  

\bibitem{Dasgupta}
  M.~Dasgupta, L.~Magnea and G.~P.~Salam,
  ``Non-perturbative QCD effects in jets at hadron colliders,''
  JHEP {\bf 0802} (2008) 055.

\bibitem{Salam:2009jx}
  G.~P.~Salam,
  ``Towards Jetography,''
  Eur.\ Phys.\ J.\ C {\bf 67} (2010) 637.


\bibitem{Catani:1991hj}
  S.~Catani, Y.~L.~Dokshitzer, M.~Olsson, G.~Turnock and B.~R.~Webber,
  ``New clustering algorithm for multi-jet cross sections in $\ee$ annihilation,''
  Phys.\ Lett.\ B {\bf 269} (1991) 432.

\bibitem{Catani:1993hr}
  S.~Catani, Y.~L.~Dokshitzer, M.~H.~Seymour and B.~R.~Webber,
  ``Longitudinally invariant $k_t$ clustering algorithms for hadron hadron collisions,''
  Nucl.\ Phys.\ B {\bf 406} (1993) 187.

\bibitem{Ellis:1993tq}
  S.~D.~Ellis and D.~E.~Soper,
  ``Successive combination jet algorithm for hadron collisions,''
  Phys.\ Rev.\ D {\bf 48} (1993) 3160.

\bibitem{Cacciari:2008gp}
  M.~Cacciari, G.~P.~Salam and G.~Soyez,
  ``The anti-$k_t$ jet clustering algorithm,''
  JHEP {\bf 0804} (2008) 063.

  \bibitem{CamOrig}
  Y.~L.~Dokshitzer, G.~D.~Leder, S.~Moretti and B.~R.~Webber,
  ``Better jet clustering algorithms,''
  JHEP {\bf 9708}, 001 (1997).

\bibitem{CamWobisch}
  M.~Wobisch and T.~Wengler,
   ``Hadronization corrections to jet cross sections in deep-inelastic
  scattering,''
  arXiv:hep-ph/9907280;
  M.~Wobisch,
   ``Measurement and QCD analysis of jet cross sections in deep-inelastic
  positron proton collisions at s**(1/2) = 300-GeV,''
DESY-THESIS-2000-049.
  
  
\bibitem{Gerwick:2012fw}
  E.~Gerwick, S.~Schumann, B.~Gripaios and B.~Webber,
  ``QCD Jet Rates with the Inclusive Generalized $k_t$ Algorithms,''
  JHEP {\bf 1304} (2013) 089.



\bibitem{Konishi:1979cb}
  K.~Konishi, A.~Ukawa and G.~Veneziano,
  ``Jet Calculus: A Simple Algorithm for Resolving QCD Jets,''
  Nucl.\ Phys.\ B {\bf 157} (1979) 45.

\bibitem{Dokshitzer:1991wu}
  Y.~L.~Dokshitzer, V.~A.~Khoze, A.~H.~Mueller and S.~I.~Troian,
  ``Basics of perturbative QCD,''
  Gif-sur-Yvette, France: Ed. Frontieres (1991) 274 p. 

\bibitem{Ellis:1991qj}
  R.~K.~Ellis, W.~J.~Stirling and B.~R.~Webber,
  ``QCD and collider physics,''
  Camb.\ Monogr.\ Part.\ Phys.\ Nucl.\ Phys.\ Cosmol.\  {\bf 8} (1996) 1.
  
 \bibitem{Herwig}
  M.~Bahr, S.~Gieseke, M.~A.~Gigg, D.~Grellscheid, K.~Hamilton, O.~Latunde-Dada, S.~Platzer and P.~Richardson {\it et al.},
  ``Herwig++ Physics and Manual,''
  Eur.\ Phys.\ J.\ C {\bf 58} (2008) 639.
  
\bibitem{Pythia8}  
  T.~Sjostrand, S.~Mrenna and P.~Z.~Skands,
  ``A Brief Introduction to PYTHIA 8.1,''
  Comput.\ Phys.\ Commun.\  {\bf 178} (2008) 852;
  T.~Sjostrand, S.~Ask, J.~R.~Christiansen, R.~Corke, N.~Desai, P.~Ilten, S.~Mrenna and S.~Prestel {\it et al.},
  ``An Introduction to PYTHIA 8.2,''
  arXiv:1410.3012 [hep-ph].

\bibitem{Cteq}
  J.~Pumplin, D.~R.~Stump, J.~Huston, H.~L.~Lai, P.~M.~Nadolsky and W.~K.~Tung,
  ``New generation of parton distributions with uncertainties from global QCD analysis,''
  JHEP {\bf 0207} (2002) 012.

\bibitem{MRST}
  A.~Sherstnev and R.~S.~Thorne,
  ``Different PDF approximations useful for LO Monte Carlo generators,''
  arXiv:0807.2132 [hep-ph].


  \bibitem{Delphes}
  S.~Ovyn, X.~Rouby and V.~Lemaitre,
  ``DELPHES, a framework for fast simulation of a generic collider experiment,''
  arXiv:0903.2225 [hep-ph].

 \bibitem{Fastjet}
  M.~Cacciari, G.~P.~Salam and G.~Soyez,
  ``FastJet user manual,''
  Eur.\ Phys.\ J.\ C {\bf 72} (2012) 1896;
  M.~Cacciari and G.~P.~Salam,
  ``Dispelling the $N^{3}$ myth for the $k_t$ jet-finder,''
  Phys.\ Lett.\ B\ {\bf 641} (2006) 57.
  
  \bibitem{Pythia6}
  T.~Sjostrand, S.~Mrenna and P.~Z.~Skands,
  ``PYTHIA 6.4 Physics and Manual,''
  JHEP {\bf 0605}, 026 (2006).

\bibitem{Bertolini:2013iqa} 
  D.~Bertolini, T.~Chan and J.~Thaler,
  ``Jet Observables Without Jet Algorithms,''
  JHEP {\bf 1404} (2014) 013.

\bibitem{Larkoski:2014uqa} 
  A.~J.~Larkoski, D.~Neill and J.~Thaler,
  ``Jet Shapes with the Broadening Axis,''
  JHEP {\bf 1404} (2014) 017.

\bibitem{TMVA}
  A.~Hocker, J.~Stelzer, F.~Tegenfeldt, H.~Voss, K.~Voss, A.~Christov, S.~Henrot-Versille and M.~Jachowski {\it et al.},
  ``TMVA - Toolkit for Multivariate Data Analysis,''
  PoS ACAT {\bf } (2007) 040
  [physics/0703039 [PHYSICS]];
  P.~Speckmayer, A.~Hocker, J.~Stelzer and H.~Voss,
  ``The toolkit for multivariate data analysis, TMVA 4,''
  J.\ Phys.\ Conf.\ Ser.\  {\bf 219} (2010) 032057;
{\tt http://tmva.sourceforge.net} .

 
\bibitem{Bolzoni} 
  P.~Bolzoni, B.~A.~Kniehl and A.~V.~Kotikov,
  ``Gluon and quark jet multiplicities at N$^3$LO+NNLL,''
  Phys.\ Rev.\ Lett.\  {\bf 109} (2012) 242002;
  P.~Bolzoni, B.~A.~Kniehl and A.~V.~Kotikov,
  ``Average gluon and quark jet multiplicities at higher orders,''
  Nucl.\ Phys.\ B {\bf 875} (2013) 18, and references therein.

\end{thebibliography}
 \end{document}